\documentclass[apj]{emulateapj}
\usepackage[hyperindex,breaklinks]{hyperref}

\def\spose#1{\hbox to 0pt{#1\hss}}
\def\simlt{\mathrel{\spose{\lower 3pt\hbox{$\mathchar"218$}}
     \raise 2.0pt\hbox{$\mathchar"13C$}}}
\def\simgt{\mathrel{\spose{\lower 3pt\hbox{$\mathchar"218$}}
     \raise 2.0pt\hbox{$\mathchar"13E$}}}
\slugcomment{Accepted for publication in ApJ, June 23 2007}
\font\smcap=cmcsc10

\def\kms{\,km~s$^{-1}$}
\def\degree{$^\circ$}
\def\arcsec{$''$}
\def\arcmin{$'$}
\def\nai{Na\,{\smcap i}}

\def\ddo{DDO51}

\def\vio{$(V-I)_0$}
\def\ivi{($I,\,V-I$)}

\def\fehp{$\rm[Fe/H]$}

\def\pg{$P_{\rm giant}$}
\def\pd{$P_{\rm dwarf}$}
\def\vhel{$v_{\rm hel}$}

\def\ol{$\langle L_i\rangle$}
\def\olnv{$\langle L_i\rangle_{\rm v\!\!\!/}$}
\def\mv{$\langle v\rangle$}
\def\mvsph{$\langle v\rangle^{\rm sph}$}
\def\sigvsph{$\sigma^{\rm sph}_{\rm v}$}
\def\mvsb{$\langle v\rangle^{\rm sub}$}
\def\sigvsb{$\sigma^{\rm sub}_v$}
\def\Gsph{$G^{\rm sph}(v)$}

\shorttitle{Substructure along M31's SE Minor Axis}
\shortauthors{Gilbert et~al.}

\begin{document}

\title{Stellar Kinematics in the Complicated Inner Spheroid of M31: Discovery of Substructure Along the Southeastern Minor Axis and its Relationship to the Giant Southern Stream}

\author{
Karoline~M.~Gilbert\altaffilmark{1},
Mark Fardal\altaffilmark{2},
Jasonjot~S.~Kalirai\altaffilmark{1,3},
Puragra~Guhathakurta\altaffilmark{1},
Marla~C.~Geha\altaffilmark{4,5},
Jedidah Isler\altaffilmark{6,7},
Steven~R.~Majewski\altaffilmark{8},
James~C.~Ostheimer\altaffilmark{8},
Richard~J.~Patterson\altaffilmark{8},
David~B.~Reitzel\altaffilmark{9}, 
Evan Kirby\altaffilmark{1},
and Michael~C.~Cooper\altaffilmark{10}
}

\email{
kgilbert@ucolick.org,
fardal@fcrao1.astro.umass.edu,
jkalirai@ucolick.org,
raja@ucolick.org,
marla.geha@nrc-cnrc.gc.ca,
jcisler@ucolick.org,
srm4n@virginia.edu,
jostheim@alumni.virginia.edu,
rjp0i@virginia.edu,
reitzel@astro.ucla.edu,
ekirby@ucolick.org,
cooper@astron.berkeley.edu
}

\altaffiltext{1}{UCO/Lick Observatory, Department of Astronomy \&
Astrophysics, University of California Santa Cruz, 1156 High Street, Santa
Cruz, California 95064.}
\altaffiltext{2}{Department of Astronomy, University of Massachusetts, Amherst, Massachusetts 01003.}
\altaffiltext{3}{Hubble Fellow.}
\altaffiltext{4}{NRC Herzberg Institute of Astrophysics, 5701 West Saanich Road, Victoria, British Columbia, Canada V9E 2E7.}
\altaffiltext{5}{Plaskett Fellow.}
\altaffiltext{6}{Fisk University/Vanderbilt University, Nashville, Tennessee 37325.} 
\altaffiltext{7}{Visiting Student, University of California Santa Cruz.}
\altaffiltext{8}{Department of Astronomy, University of Virginia, PO
Box~400325, Charlottesville, VA 22904-4325.}
\altaffiltext{9}{Department of Physics \& Astronomy, Knudsen Hall, University
of California, Los Angeles, California 90095.}
\altaffiltext{10}{Department of Astronomy, Campbell Hall, University of
California, Berkeley, California 94720.}
\setcounter{footnote}{10}

\begin{abstract}
We present the discovery of a kinematically-cold stellar population along 
the southeastern minor axis of the Andromeda galaxy (M31) that is likely the forward
continuation of M31's giant southern stream.  This
discovery was made in the course of an on-going spectroscopic survey of red giant
branch (RGB) stars in M31 using the DEIMOS instrument on the Keck~II 10-m
telescope\footnote{Data presented herein were obtained at the W.\ M.\ Keck
Observatory, which is operated as a scientific partnership among the
California Institute of Technology, the University of California and the
National Aeronautics and Space Administration.  The Observatory was made
possible by the generous financial support of the W.\ M.\ Keck Foundation.}.
Stellar kinematics are investigated in eight fields
located 9\,--\,30~kpc from M31's center (in projection).  
A likelihood method based on photometric and spectroscopic diagnostics is
used to isolate confirmed M31 RGB stars from foreground
Milky Way dwarf stars: for the first time, this is done {\it without\/}
using radial velocity as a selection criterion, allowing an unbiased
study of M31's stellar kinematics.
The radial velocity distribution of the 1013 M31 RGB stars shows evidence
for the presence of two components.
The broad (hot) component has a velocity dispersion of \sigvsph=129~\kms\ and 
presumably represents M31's virialized spheroid.   
A significant fraction (19\%) of the population is in a narrow (cold)
component centered near M31's systemic velocity
with a velocity dispersion that decreases with 
increasing radial distance, from \sigvsb$=55.5$~\kms\ at $R_{\rm proj}=12$ kpc 
to \sigvsb$=10.6$~\kms\ 
(an \textit{intrinsic} velocity dispersion of 9.5~\kms\ after 
accounting for velocity measurement error) 
at $R_{\rm proj}=18$ kpc.   
The spatial and velocity distribution of the cold component 
matches that of the ``Southeast shelf'' predicted by the \citet{far07} 
orbital model of the progenitor of the giant southern stream.
The metallicity distribution of the cold component matches that
of the giant southern stream, but is about 0.2~dex more metal
rich on average than that of the hot spheroidal component.
We discuss the 
implications of our discovery on the interpretation of the
intermediate-age spheroid population found in this region in recent ultra-deep
{\it HST\/} imaging studies.

\end{abstract}



\section{Introduction}\label{sec:intro}

\setcounter{footnote}{11}

In the current paradigm of hierarchical galaxy formation, massive galaxies are built up 
through a series of major and minor merger events \citep{sea78,whi78}.  
Observations of galaxies at high redshift show that merging systems 
are common \citep[e.g.,][]{abr96,lef00,con03,lot07}, and large-scale 
simulations of galaxy formation in
a cosmological context have successfully reproduced many of the observed
properties of galaxies and galaxy clusters \citep[e.g.,][]{spr05,cro06}.  

A consequence of hierarchical galaxy formation is that galactic stellar 
halos should be at least partially composed of the tidal debris from past
accretion events.  Numerical simulations and semi-analytic models of 
stellar halo formation have made great strides toward understanding 
the properties of halos that are built up through tidal stripping of 
merging systems \citep[e.g.,][]{joh96,joh98,hel99,hel00,bul01,bul05}.  
Detailed comparisons between observations and simulations are needed
to determine the fraction
of stellar halos that are made up of tidal debris and to better
understand the formation of galaxies in general. 
 
Recent discoveries of tidal streams in the stellar halos of 
the Milky Way (MW) and Andromeda (M31) galaxies
are providing the most direct and detailed observational constraints on theories of stellar halo 
formation.  Among the most prominent of these substructures are the Sagittarius 
stream \citep{iba94,maj03,new03}, the 
Monoceros stream \citep{yan03,roc03}, and the Magellanic stream \citep{mat74} 
in the MW, and the giant southern stream
\citep[GSS;][]{iba01} in M31.  Additional substructures have been identified in 
M31 that are also likely remnants of past mergers, such as the Northeast shelf 
\citep{fer02,iba04,far06}, a secondary cold component in the same physical location as the giant
southern stream \citep{kal06a}, and the
various substructures identified with the disk of M31 \citep{iba05}. 
Tidal disruption has also been observed in M31's satellite galaxies M32 and NGC 205
\citep{cho02}.

In addition to providing insight into theoretical models of stellar halo
formation, the observed properties of tidal streams can be used to constrain the the galactic 
potential in which they are found if sufficient 
phase-space information is available
\citep[e.g.,][]{joh99,joh02,pen06}.  Several attempts have been made to model the 
mass distribution of the MW using observed substructure, 
most of which have focused on the orbital properties of the Sagittarius stream 
\citep[e.g.,][]{iba01a,hel04,mar-del04,joh05,law05,fel06}. 

The GSS 
has been the focus of detailed modeling in M31, spurred on by recent 
observations.  Imaging and photometry 
have revealed the physical extent of the GSS
\citep{fer02,mccon03,fer06} and provided line-of-sight distances
at various points along it \citep{mccon03}, while spectroscopy
has yielded the mean line-of-sight velocity and velocity
dispersion of stream stars as a function of position
\citep{iba04,guh06,kal06a}.  The availability of this phase-space 
information has motivated
several groups to model the orbit of the progenitor of the GSS
\citep{iba04,fon06,far06,far07}.  However, \citet{far06} concluded 
that the degree to which the GSS can be used to constrain M31's mass 
distribution is limited by 
the current measurement uncertainties in the distance to the stream 
and the lack of a clearly identified compact stellar concentration that might
correspond to the dense remnant core of the stream's progenitor galaxy.
Further observational
constraints on the orbit of the progenitor of the GSS, such 
as the identification of tidal debris from other pericentric
passages, are needed to make progress.

Towards this end, \citet[][hereafter F07]{far07} show that several of the observed
features in M31 can
be explained as the forward continuation 
of the GSS.  Their model makes predictions that
can be tested by observations, including the stellar velocity distributions
in the Northeast
and Western shelves and the presence of a weaker shelf on the 
eastern side of the galaxy. This last shelf is expected to be most 
easily visible near the southeastern minor axis of M31\footnote{Although this
shell feature is expected to span an $\sim 180$\degree\ range in position angle, covering
the eastern half of the galaxy, it is expected to be most easily observable in the southeast
due to overlap with the much denser Northeast shelf and M31's disk elsewhere
(see Fig.~\ref{fig:xieta_sim}).}.  

This paper presents new substructure along M31's southeastern minor axis at
the expected location of the F07 Southeast shelf and displaying the distinct
kinematic profile predicted by their orbital model.  The substructure
was discovered in the course of an on-going Keck/DEIMOS spectroscopic study of the
dynamics and metallicity of RGB stars in the inner spheroid and outer halo of
M31 \citep[see][and references therein]{gil06}.  The portion of the 
inner spheroid studied here appeared to be relatively undisturbed in 
earlier star count maps \citep{fer02} and radial velocity surveys 
\citep{rei02, kal06a}. 
The photometric and spectroscopic data used in this analysis are described in 
\S\,\ref{sec:data}.  The criteria for selection of 
M31 RGB stars are discussed in \S\,\ref{sec:sample}.
The kinematics of the RGB population 
(first the combined data set and then the individual fields)
are characterized in \S\,\ref{sec:kin}.
The spatial trends and general properties of the dynamically hot
spheroid and cold substructure populations are discussed in
\S\,\ref{sec:veldisp} and \S\,\ref{sec:coldpop}, respectively.
The physical origin of the cold substructure is 
explored in 
\S\,\ref{sec:origin}, including its likely relation to M31's GSS.
The relevance
of the newly discovered cold substructure to
the \citet{bro03} discovery of an intermediate-age population in the spheroid of M31 
is discussed \S\,\ref{sec:intm_age}.
The main conclusions of the paper are summarized in \S\,\ref{sec:concl}.

\section{Data}\label{sec:data}
\begin{figure}
\epsscale{0.85}
\plotone{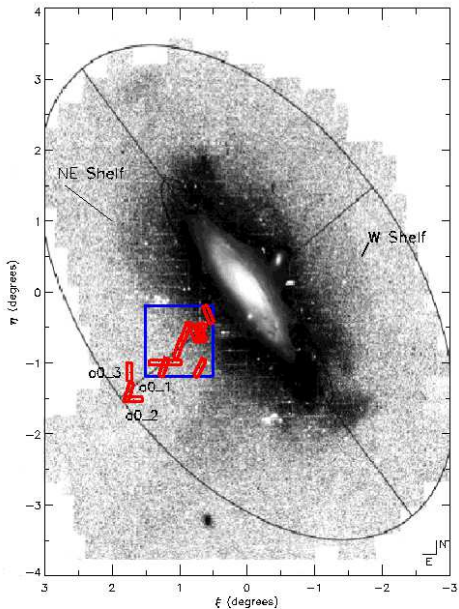}
\caption{
Sky positions of the fields discussed in this paper.  
The blue square represents the position and area of the CFHT/MegaCam image 
(Fig.~\ref{fig:cfht}). 
The red rectangles approximate 
the size and position angle of the DEIMOS spectroscopic slitmasks.  The three
masks nearest the outer ellipse are in field a0.
The remaining masks are identified in Figure~\ref{fig:cfht}. 
  The star count map comes from \citet{iba05}, and is in 
standard M31-centric coordinates ($\xi, \eta$).  The outer ellipse represents 
a 55 kpc radius along the major axis, with a flattening of 3:5.  
The major and minor axes of M31 are indicated by straight lines.  
The giant southern 
stream is the obvious overdensity of stars south of M31's center.   
}
\label{fig:roadmap}
\end{figure}

\begin{figure}
\plotone{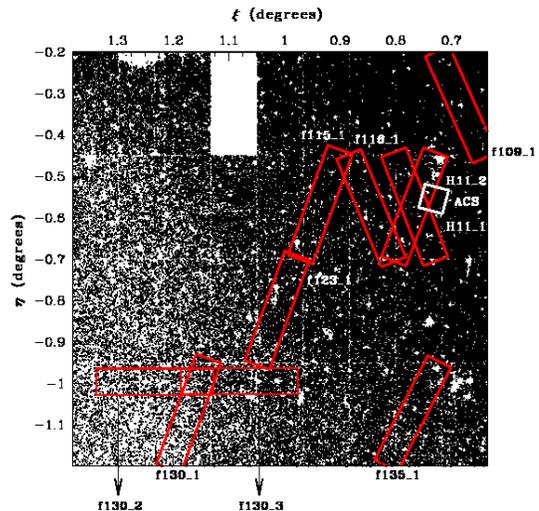}
\caption{
Starcount map derived from CFHT/MegaCam photometry in a single pointing
with the 36-CCD mosaic (\S\,\ref{sec:phot}). The orientation of this map is 
the same as Figure~\ref{fig:roadmap}.
The size and positions of the masks designed from the CFHT/MegaCam photometry 
are
shown as red rectangles. The white square shows the position and approximate 
orientation of the \citet{bro03} deep HST/ACS observations (\S\,\ref{sec:intm_age}). 
The three a0 masks were 
based on photometry from the KPNO 4-m telescope and are to the southeast (bottom left), beyond the limit of 
this image.  There is an apparent edge in the density of star counts in the image, 
running from the upper left to the lower right and passing through field f123; 
this feature will be discussed in \S\,\ref{sec:gss}.  
}
\label{fig:cfht}
\end{figure}

The data set discussed in this paper is drawn from photometry and spectroscopy of several fields 
on/near the southeastern minor axis of M31.  The locations of the 
fields are shown in Figures~\ref{fig:roadmap} and \ref{fig:cfht}.  They span a 
range of projected radial 
distances from M31's center of $R_{\rm proj}\sim 9$ to 30~kpc (Table~\ref{tab:masks}).   
A brief explanation of the data sets and reduction is included below.  A more
detailed discussion of the  
observational strategy and data reduction methods employed in our M31 survey can be 
found in \citet{guh06}, \citet{kal06a,kal06b}, and \citet{gil06}.  

\subsection{Photometry}\label{sec:phot}

Photometry and astrometry for the majority of the fields analysed in this paper
were derived from MegaCam images in the 
$g'$ and $i'$ bands obtained with
the 3.6-m Canada-France-Hawaii Telescope (CFHT)\footnote{MegaPrime/MegaCam is
a joint project of CFHT and CEA/DAPNIA, at the Canada-France-Hawaii Telescope
which is operated by the National Research Council of Canada, the Institut
National des Science de l'Univers of the Centre National de la Recherche
Scientifique of France, and the University of Hawaii.}.  
The program SExtractor \citep{ber96} was used for object 
detection, photometry, and morphological classification (via the {\tt stellarity}
parameter).  
The instrumental $g'$ and $i'$ magnitudes were 
transformed to Johnson-Cousins $V$ and $I$ magnitudes based on observations
of Landolt photometric standard stars \citep{kal06a}.

Photometry and astrometry for a0, the outermost field discussed in this paper, 
were derived by \citet{ost02} from images obtained with the Mosaic camera 
on the Kitt Peak National Observatory (KPNO)\footnote{Kitt Peak National 
Observatory of the National Optical Astronomy Observatory is operated by the 
Association of Universities for Research in Astronomy, Inc., under cooperative 
agreement with the National Science Foundation.} 4-m telescope 
in the Washington System $M$ and
$T_2$ bands and the intermediate-width \ddo\ band.  This combination 
of filters allows photometric selection of stars that are likely to be
M31 red giant branch (RGB) stars rather than MW dwarf stars \citep[e.g.,][]{pal03, maj05}.  
The \ddo\ filter
is centered at a wavelength of 5150~\AA\ with a width of $\sim 100$~\AA, and 
includes the surface-gravity sensitive Mg\,$b$ and MgH stellar absorption
features which are strong in dwarf stars but weak in RGB stars.  Based on a 
star's position in the ($M-DDO51$) versus ($M-T_2$) color-color diagram, it 
is assigned a probability of being an M31 RGB star. 
Johnson-Cousins $V$ and $I$ magnitudes were derived from the $M$ and 
$T_2$ magnitudes using the photometric transformation relations in \citet{maj00}.
Use of \ddo\ photometry to screen for likely M31 RGB stars increases 
the efficiency of the spectroscopic observations,   
suppressing the number of selected MW dwarf stars by a factor of $\approx 3$ 
in a0 (Guhathakurta et al. 2007, in prep).

\subsection{Spectroscopy}
\subsubsection{Slitmask Design and Observations}\label{sec:slit_design}

Objects in fields covered by the CFHT/MegaCam images were selected for 
Keck/DEIMOS spectroscopy based on $I$ magnitude and
the SExtractor morphological criterion {\tt stellarity} \citep{kal06a}.  
Objects in field a0 were selected on the basis of $I$ magnitude and  
morphological criteria (DAOPHOT parameters {\tt chi} and {\tt sharp}), prioritized according to their  
probability of being an M31 RGB star (based on $M$, $T_2$, and 
\ddo\ photometry as described in \S\,\ref{sec:phot}).  Pre-selection of likely M31 RGB 
stars is vital for efficient study of the sparse outer parts of the M31 halo.  The 
inner fields (such as the fields drawn from the CFHT/MegaCam images) contain
a relatively high 
surface density of M31 RGB stars, so the RGB to MW 
dwarf star ratio is high even without \ddo-based pre-selection of RGB candidates.

For the purposes of most of the analysis in this paper, data are classified according
to fields, rather than spectroscopic masks.  In general, a ``field'' refers to 
the area covered by a single CFHT/MegaCam CCD ($\sim 15$\arcmin$\times$6\farcm5, Fig.~\ref{fig:cfht}); there 
can be one or more overlapping 
DEIMOS masks (16\arcmin$\times$4\arcmin) in a single field.  For example,
masks H11\_1 and H11\_2 are both part of field H11.
There are two exceptions to this field/mask scheme: (1) the a0 field refers to a single Mosaic pointing,
which covers a 35\arcmin$\times$35\arcmin\ area, and (2)
two of the f130 masks each straddle a couple of adjacent MegaCam CCDs,
but since they were chosen to overlap with the f130\_1 mask, they have 
been labelled f130\_2 and f130\_3. 

Fields were observed using the Keck~II telescope and the DEIMOS instrument with 
the 1200~line~mm$^{-1}$ grating.  Most of the fields were observed in Fall~2005 [f109, f115, f116,
f123, f130 (1 mask), and f135].  The a0 masks were observed in Fall~2002 
and 2004, the H11 masks in 2004, and the last two of the three f130 
masks in Fall~2006
(Table~\ref{tab:masks}).
The central wavelength for most masks was $\rm\lambda7800~\AA$, yielding a 
spectral coverage of approximately $\lambda\lambda6450$--$\rm9150~\AA$.  The only exceptions were
the a0\_1 and a0\_2 masks observed in 2002, which had a central wavelength of 
$\rm\lambda8550~\AA$
and a spectral coverage of approximately $\lambda\lambda7200$--$\rm9900~\AA$.    
The 1200~line~mm$^{-1}$ grating has a dispersion of $\rm0.33~\AA$~pix$^{-1}$;
 the scale in the spatial direction is 0$\farcs$12~pix$^{-1}$, and the effective scale in
the dispersion direction is 0$\farcs$21~pix$^{-1}$.  Slits had a width of
1\arcsec.  The spectral resolution for a star observed in typical $0.8''$ 
seeing conditions is about 1.3~\AA\ FWHM.  Each mask was observed for a 
total of 1~hour,
with the exception of field f109 which was observed for 3~hours. 

\begin{deluxetable*}{rcrrrlrr}
\tablecolumns{8}
\tablewidth{0pc}
\tablecaption{Details of Spectroscopic Observations and Basic Results.}
\tablehead{\multicolumn{1}{c}{Mask} & \multicolumn{1}{c}{Projected}          & \multicolumn{2}{c}{Pointing center:} &
  \multicolumn{1}{c}{PA} & \multicolumn{1}{c}{Date of} & 
  \multicolumn{1}{c}{\#\ Sci.}  & \multicolumn{1}{c}{\#\ of M31} \\ 
       & \multicolumn{1}{c}{Radius}     & \multicolumn{1}{c}{$\alpha_{\rm J2000}$} &
\multicolumn{1}{c}{$\delta_{\rm J2000}$} & \multicolumn{1}{c}{($^\circ$E of N)}
& \multicolumn{1}{c}{Obs. (UT)} & \multicolumn{1}{c}{targets\tablenotemark{a}} &
\multicolumn{1}{c}{Stars\tablenotemark{a,\rm b}} \\
     & \multicolumn{1}{c}{(kpc)} & \multicolumn{1}{c}{($\rm^h$:$\rm^m$:$\rm^s$)} &
\multicolumn{1}{c}{($^\circ$:$'$:$''$)} &  &} 
\startdata
f109\_1  & 9 & 00:45:46.75 &  +40:56:53.8 & 23.90  &   2005 Aug 29  & 208 & 169 \\
H11\_1   & 12 & 00:46:21.02 & +40:41:31.3 & 21.0     & 2004 Sep 20 & 139 & 89 \\ 
H11\_2   & 12 & 00:46:21.02 & +40:41:31.3 & $-21.0$  & 2004 Sep 20 & 138 & 88 \\
f116\_1  & 13 &   00:46:54.53 &   +40:41:29.5  & 22.60     &  2005 Aug 28 & 199 & 149 \\
f115\_1  & 15 & 00:47:32.71 &   +40:42:00.9  & $-20.0$     &  2005 Aug 28  & 191 & 114 \\
f123\_1  & 18 &  00:48:05.57 &  +40:27:16.3 & $-20.0$   & 2005 Aug 28 & 171 & 104 \\
f135\_1  & 18 &  00:46:24.88  &   +40:11:35.5   & $-27.0$     &  2005 Aug 29 & 146 & 99\\
f130\_1  & 22 &  00:49:11.97 &   +40:11:45.3  & $-20.0$    &  2005 Aug 28 & 108 & 52 \\
f130\_2  & 23 &  00:49:37.49 &  +40:16:07.0  & 90.0    &  2006 Nov 21 & 115 & 43 \\
f130\_3  & 20 &  00:48:34.59 & +40:16:07.0 & 90.0    &  2006 Nov 22 & 124 & 41 \\
a0\_1    & 31 & 00:51:51.32 & +39:50:21.4 & $-17.9$ &  2002 Aug 16 & 89 & 25 \\
a0\_2    & 31 & 00:51:29.59 & +39:44:00.8 & 90.0    &  2002 Oct 12 & 89 & 32 \\
a0\_3    & 29 & 00:51:50.46 & +40:07:00.9 & 0.0     &  2004 Jun 17 & 90 & 26 \\
\enddata
\tablenotetext{a}{A number of targets were observed on two different masks.
Therefore, the total number of unique science targets/M31 RGB stars in fields H11, f130 
and a0 is less than the reported number.  There are 18 M31 RGB
stars with duplicate observations (2 in H11, 8 in f130, and 8 in a0), thus the
total number of unique M31 RGB stars is 1013.}
\tablenotetext{b}{The number of M31 RGB stars is defined as the number of stars
that are identified as secure and marginal M31 RGB stars by the \citet{gil06}
diagnostic method, wihout the use of the radial velocity diagnostic (\olnv$>0$, \S\,\ref{sec:sample}).}
\label{tab:masks}
\end{deluxetable*}


%
\subsubsection{Spectroscopic Data Reduction}\label{sec:dataredux}

The spectra were reduced and analyzed using a modified version of the {\tt spec2d} 
and {\tt spec1d} software developed by the DEEP2 team at the University of 
California, Berkeley\footnote{{\tt
http://astron.berkeley.edu/$\sim$cooper/deep/spec2d/primer.html},
\newline\indent
{\tt http://astron.berkeley.edu/$\sim$cooper/deep/spec1d/primer.html}
}; these
routines perform standard spectral reduction steps, including flat-fielding, 
night-sky emission line removal, and extraction of the two-dimensional spectra.
Reduced one-dimensional spectra are cross-correlated with a library of template
stellar spectra to determine the redshift of the object.  Each 
spectrum was visually inspected and assigned a quality code
based on the number and quality of absorption lines.  Spectra with at least
two spectral features (even if one of them is marginal) are 
considered to have secure redshift measurements.  A heliocentric correction 
is applied to the measured radial velocities based on the sky position of the
mask and the date and time of the observation.  
The heliocentric velocities are not corrected for the changing
component of solar motion across our fields; our innermost and outermost 
fields are separated by 1.5\degree\ along the southeastern minor axis, 
which corresponds to only a 1.6~\kms\ velocity change.

Spectra in field a0 were reduced using the original reduction techniques briefly 
outlined above.  Detailed discussions of the spectral reduction techniques, 
quality determination, and S/N measurements used in our survey can be found in
\citet{guh06} and \citet{gil06}; the typical velocity error in this field 
is 15~\kms.  Our cross-correlation procedure has since been improved and 
is described below.  An offset of $+20$~\kms\ has been applied to the 
a0 data to make them consistent with the results of this new 
cross-correlation procedure.

Spectra in the remainder of the fields were reduced using several
improvements to the reduction pipeline.  A greater number of stellar
templates are used for the spectral cross-correlation, and the template
library has been expanded to include spectral templates from the Keck II
telescope's Echellete Spectrograph and Imager (ESI) and DEIMOS in addition to
the existing Sloan Digital Sky Survey (SDSS) spectral templates.  The ESI and
DEIMOS templates were included because they more closely match the resolution
of the observed spectra.  The position of the atmospheric A-band in the
observed spectrum is used to correct the observed radial velocity for
imperfect centering of the star in the slit \citep[Simon \& Geha 2007, in
prep;][]{soh07}.  The improvement to the spectral templates and the A-band
correction allows us to reduce our velocity measurement errors relative to
our previous reductions.  The median velocity error for the data presented in 
this paper is 4.6~\kms, estimated from the cross-correlation output routine
and confirmed by repeat measurements of individual stars on overlapping DEIMOS
masks. 

Spectroscopic data in fields H11 and a0 have been presented in previous 
papers \citep{kal06a,bro03,bro06}.  
The radial velocity sample for H11 presented in this paper contains
$\approx50$\% more M31 RGB stars than the previously published sample,
due to the recent recovery of spectra from two CCDs that were not included in
these earlier papers and the
improvements in the data reduction process described above.

\section{Selecting a Sample of M31 Red Giants}\label{sec:sample}
The largest source of contaminants in our spectroscopic survey are foreground
MW dwarf stars.  Background galaxies are easy to identify and remove
from the sample on the basis of their spectra and redshifts.  However, the 
radial velocity distribution of MW dwarf stars overlaps that of M31 RGB stars, 
making identifying individual stars as M31 red giants or MW dwarfs problematic. We use the diagnostic method detailed in \citet{gil06} to separate M31 RGB
stars from MW dwarf star contaminants.  The method uses empirical probability
distribution functions to estimate the likelihood a given star is an M31 red 
giant based on five photometric and spectroscopic diagnostics: 
\begin{itemize}
\item The radial velocity of the star.
\item 
Photometry in the 
  $M$, $T_2$, and (surface-gravity sensitive) \ddo\ bands
\item The measured equivalent width of the \nai\ doublet at 8190\AA\ combined with the \vio\ color of the star.
\item The position of the star in an \ivi\ color magnitude diagram with respect to theoretical RGB isochrones.
\item A comparison of the star's photometric vs. spectroscopic metallicity estimates.
\end{itemize}
The \ddo\ diagnostic is only used for field a0, which is the only field in the 
present work for which \ddo\ photometry is available 
(\S\,\ref{sec:phot}).  The likelihoods for each diagnostic are 
combined in a weighted average for each star to determine the overall 
likelihood, \ol, the star is an M31 RGB or MW dwarf star (\S\,A.1).  
Based on the overall likelihood, each star is 
identified as either a secure M31 RGB star (\ol$>0.5$, or $>3\times$ more likely to be an M31 RGB star than 
an MW dwarf) or secure MW dwarf star 
(\ol$<-0.5$), or a marginal M31 RGB star ($0<$\ol$<0.5$
) or marginal MW dwarf star ($-0.5<$\ol$<0$).

An advantage of the diagnostic method for the present analysis is 
the ability to select a sample that is chosen independently of radial velocity. 
Since radial velocity is only one of a number of diagnostics which are used to 
determine the nature of an individual star, 
the 
likelihood method
(even with the inclusion of the radial velocity diagnostic) 
presents a significant improvement over the use of velocity cuts to select samples for 
kinematical analysis by reducing the sensitivity of the 
sample to velocity.  However, by using the likelihood
method \textit{without} the radial velocity diagnostic (resulting in overall 
likelihoods \olnv), we are able to select
a sample of M31 red giants that is completely \textit{independent} of the radial velocities
of the stars. 

\begin{figure}
\plotone{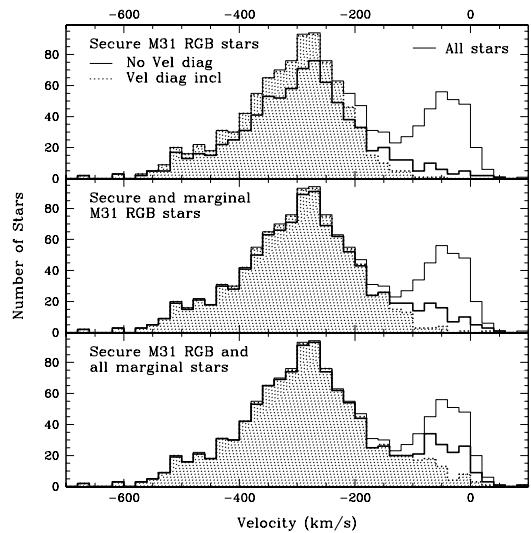}
\caption{
The radial velocity distributions of samples with (\textit{shaded/dotted histograms}) 
and without (\textit{thick open histograms}) the radial velocity diagnostic included in the 
overall likelihood calculation.  The radial velocity distribution of all stars
with successful radial velocity measurements is shown for comparison (\textit{thin open histograms}); the MW dwarf star contaminants form the peak at $v_{\rm hel}\approx -50$~\kms.  \textit{Top:} Stars designated as secure 
([\ol,\olnv]$>0.5$) M31 red giants only.  \textit{Middle:}  Stars designated as marginal and
secure M31 red giants ([\ol,\olnv]$>0$).  \textit{Bottom:} Stars 
designated as marginal MW dwarfs, marginal M31 red giants, and secure M31 red giants ([\ol,\olnv]$>-0.5$).  
The distributions are similar, but there is a deficiency of 
stars at velocities near 0~\kms\ in the sample that includes radial 
velocity as a diagnostic.   
}
\label{fig:velhist_bias}
\end{figure}

Fig. \ref{fig:velhist_bias} presents the radial velocity distribution of stars 
selected as M31 red giants based on the diagnostic method with (\textit{shaded histograms}) and 
without (\textit{thick open histograms})
radial velocity included, for multiple \ol\ and \olnv\ thresholds.  For reference, the radial velocity distribution of
all stars with successful velocity measurements is also shown 
(\textit{thin open histograms}); the MW dwarf star contaminants form the 
secondary peak centered at $v_{\rm hel}\approx -50$~\kms.  
The M31 RGB 
distributions are similar,
with the sample that includes the radial velocity diagnostic showing a systematic
deficiency of stars at radial velocities near 0~\kms.  
The effect of the 
radial velocity diagnostic on the overall likelihood (\ol\ and \olnv) distributions is discussed 
in \S\,A.1.   

The M31 RGB samples identified by their \ol\ values 
have a minimal amount of MW dwarf star contamination, but are
also kinematically biased against stars with velocities near 0~\kms\ 
(\S\,A.1).  The M31 RGB samples identified by their \olnv\ values 
have a slightly larger amount of dwarf
contamination (particularly evident in the bottom panel of Figure~\ref{fig:velhist_bias}), 
but the underlying M31 RGB population is kinematically unbiased.

The RGB sample used in this paper is defined as stars that are designated 
as secure and marginal M31 red giants by the diagnostic 
method, with the radial velocity likelihood \textit{not} included in the 
calculation of a star's overall likelihood of being an M31 RGB star (i.e., \olnv$>0$).  
The number of M31 RGB stars in each field is listed in Table~\ref{tab:masks}.  
The \olnv$>0$ threshold maximizes the completeness of the underlying, \textit{kinematically 
unbiased} M31 RGB population,
but introduces an overall MW dwarf star contamination of 5\% to the sample 
(\S\,A.2).  The contamination is largely constrained to $v_{\rm hel}>-150$~\kms\ due to the velocity distribution of MW dwarf stars, and its effect
on the measured parameters of the M31 RGB sample is quantified in 
\S\,A.3.

\section{Stellar Kinematics in M31's Southeast Minor-Axis Fields}\label{sec:kin}
The data presented in this paper span a range in projected radial distance
from the center of M31 of 9 to 30~kpc, along the southeastern minor axis.  We 
refer to this region as the ``inner spheroid,'' even though it has
traditionally been referred to as the ``halo'' of M31. 
This region departs from the classical picture of a stellar halo formed from 
observations of the 
Milky Way: M31's inner spheroid is about $10\times$ more metal-rich than the 
MW's halo \citep{mou86,dur04} and follows
a de~Vaucouleurs $r^{1/4}$ surface density profile \citep{pri94,dur04}, 
while the surface density profile of the MW halo follows an $r^{-2}$ power law.  In other words, 
M31's inner spheroid appears to be a continuation of its 
central bulge.  Recent observations have discovered an outer stellar ``halo''
in M31 which is relatively metal-poor \citep{kal06b, cha06}, has a surface density 
profile that follows an $\sim r^{-2}$ power law \citep{guh05, irw05}, and has 
been detected out to $R_{\rm proj}=165$~kpc \citep{gil06}. 

These observations
imply that the 
spheroid of M31 has two components: an inner, de~Vaucouleurs profile spheroid, and an outer, power-law profile halo.
We thus use the term \textit{inner spheroid} to distinguish the region 
$R_{\rm proj}\sim 9$\,--\,30~kpc from the canonical central 
bulge and the newly discovered stellar halo of M31.  The outer limit of this
region is well-defined; 
a break in the surface brightness profile of M31 RGB stars has been observed at 
$R_{\rm proj}\sim 20-30$~kpc \citep{guh05,irw05}, and there is observational 
evidence that the crossover between the 
predominantly metal-rich population of the inner spheroid and the 
predominantly metal-poor population of the outer halo occurs at 
$R_{\rm proj}\sim 30$~kpc \citep{kal06b}.  The inner limit of this 
region is arbitrary, as the relationship between this component and 
the central bulge of M31 is not yet clear. 

The rest of this section characterizes the line of sight velocity 
distribution of stars in the inner spheroid of M31 through
maximum-likelihood fits of Gaussians to the combined data set 
(\S\,\ref{sec:anal_comb}) and to individual fields (\S\,\ref{sec:anal_ind}). 
Gaussians provide a convenient means of fitting 
for multiple kinematical components in the data and characterizing their
mean velocity and velocity dispersion.  The true shape of the velocity
distribution of a structural component in M31 is likely to be different
from a pure Gaussian.  However, given the limited sample size and the absence
of any specific physical model, the choice of Gaussians seems appropriate. 

\subsection{Maximum-Likelihood Fits to the Velocity Distribution of the Combined Data Set}\label{sec:anal_comb}

\begin{figure}
\plotone{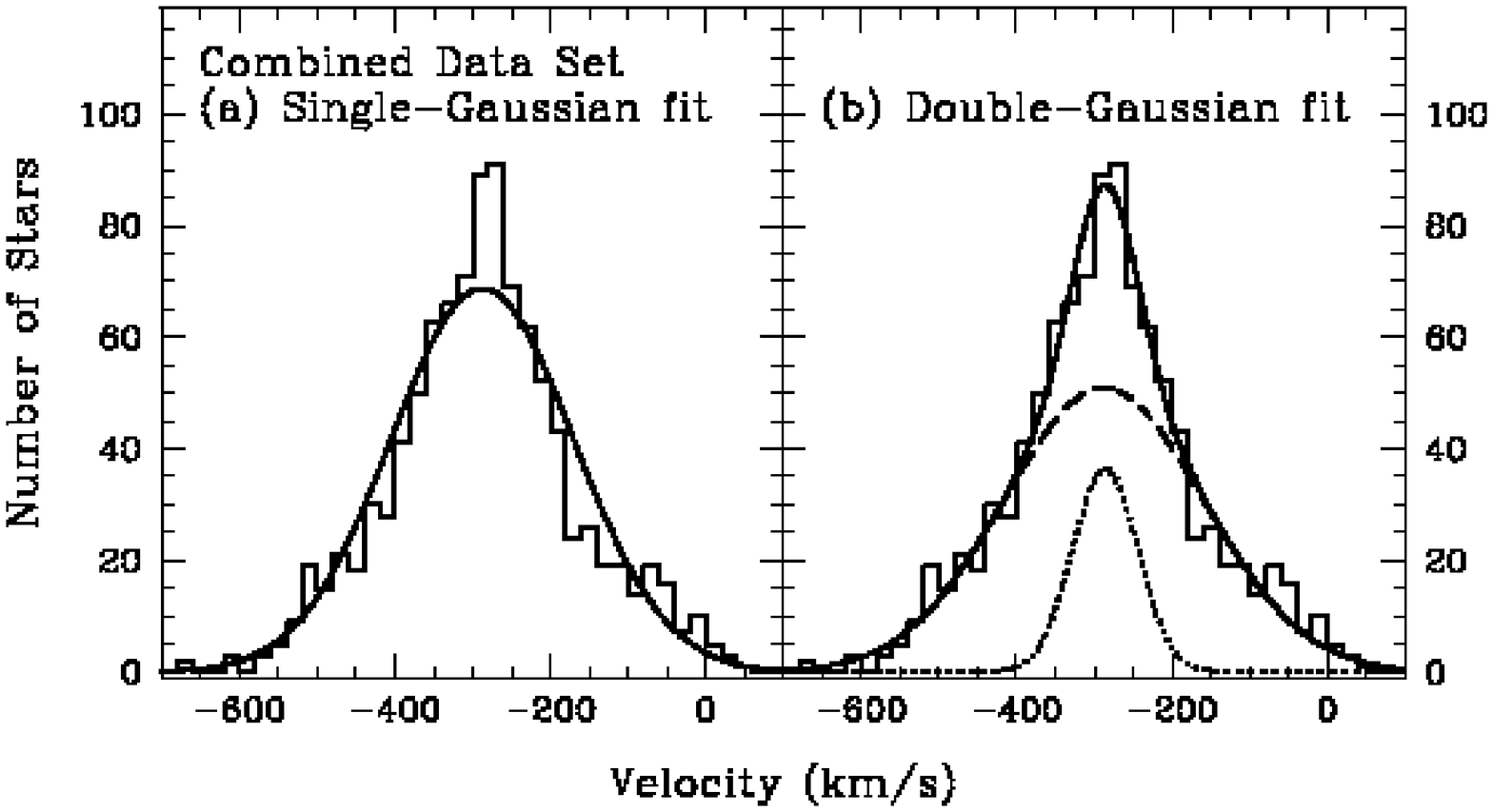}
\caption{
The radial velocity distribution of the M31 RGB inner spheroid population.  
A maximum-likelihood analysis was used to fit an analytic function to 
the distribution. (\textit{a}) The best fit single Gaussian has parameters \mvsph$=-287$~\kms 
and \sigvsph=117~\kms.  A $\chi^2_{\rm red}$ test rules out the single-Gaussian fit at a very high
confidence level.  (\textit{b}) The best constrained double-Gaussian fit 
(Table~\ref{table:rvfits}) is shown as a solid curve, with the narrow component  
and wide components displayed separately as dotted and dashed curves, respectively. 
}
\label{fig:vel_all}
\end{figure}

Figure~\ref{fig:vel_all} shows the combined radial velocity distribution of 
M31 RGB stars from all eight fields along the minor axis, ranging from 9 
to 30~kpc in projected radial distance from the center of M31.
Fits to the radial velocity distribution were made using a maximum-likelihood 
technique; the best-fit single (\textit{a}) and double (\textit{b}) 
Gaussians are displayed in Figure~\ref{fig:vel_all}. 
A reduced $\chi^2$ analysis rules out the single-Gaussian fit, as the probability
is $<\!\!<\!\!1\%$ that the observed radial velocities were drawn from the best-fit
distribution.  The observed velocity distribution is well fit
by a sum of two Gaussians (solid curve, panel \textit{b} of Fig.~\ref{fig:vel_all}), composed of a wide Gaussian (dashed curve) centered at \mvsph=
$-287.2^{+8.0}_{-7.7}$~\kms, with a width of 
\sigvsph$=128.9^{+7.7}_{-6.9}$~\kms, and a narrow Gaussian (dotted curve)  
centered at \mvsb=$-285.4^{+12.8}_{-12.4}$~\kms\ with a width of 
\sigvsb=$42.2^{+12.5}_{-14.3}$~\kms, which comprises $19^{+9}_{-8}$\% of 
the total population (quoted errors represent the 90\% confidence
limits from the maximum-likelihood analysis).  Due to the
MW dwarf star contaminants in the M31 RGB sample (\S\,\ref{sec:sample}), the true
\mvsph\ value of the wide Gaussian component is 15 to 20~\kms\ more 
negative than the best-fit value (\S\,A.3), making it consistent with 
the systemic velocity of M31 ($v_{\rm sys}=-300$~\kms). 
The kinematically hot component
corresponds to the inner spheroid of M31 (quantities related to this component are 
denoted with the subscript ``sph''), while the kinematically cold
component corresponds to substructure in the inner spheroid (denoted with
the subscript ``sub''); the discussion
of these components
and their properties will be deferred to \S\,\ref{sec:veldisp} and 
\S\,\ref{sec:coldpop}, respectively.  The wider of the two Gaussian
components in the double Gaussian fit to the combined data set is hereafter
referred to as \Gsph.

\begin{figure}
\plotone{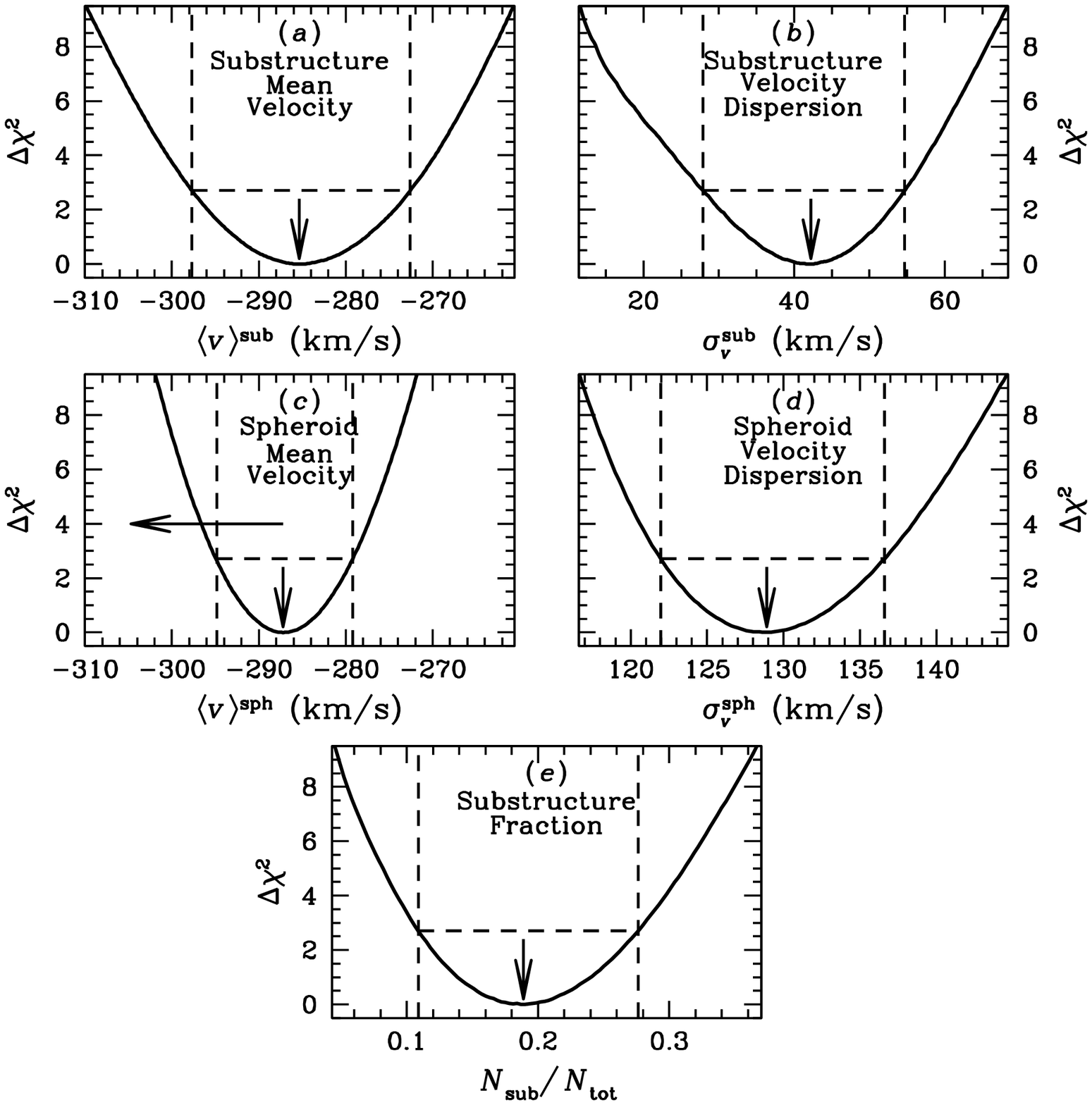}
\caption{
Results of the maximum-likelihood analysis for the double-Gaussian
fit to the combined M31 RGB inner spheroid sample.  The optimal value of 
each parameter is marked with an arrow, and the 90\% confidence limits from
the maximum-likelihood analysis are
marked with dashed lines.  The upper limit of the y-axis represents the 
99\% confidence limits.  The parameter $\Delta\chi^2\equiv \chi^2-\chi_{\rm min}^2$ 
is plotted as a function of  
(\textit{a}) mean velocity of the cold substructure component (narrow Gaussian), 
\mvsb, (\textit{b}) velocity dispersion of the cold component, \sigvsb, 
(\textit{c}) mean velocity of the M31 inner spheroid (wide Gaussian), \mvsph,
(\textit{d}) velocity dispersion of the M31 inner spheroid, \sigvsph, and
(\textit{e}) fraction of the total M31 RGB population in the cold component,
$N_{\rm sub}/N_{\rm tot}$.  
The horizontal arrow in panel (\textit{c})
represents the correction to the \mvsph\ value necessary to offset the
bias caused by MW dwarf contamination at $v_{\rm hel}>-150$~\kms\ (\S\,A.3). 
}
\label{fig:errors_all}
\end{figure}

Figure~\ref{fig:errors_all} shows error estimates from the maximum-likelihood
analysis (in the form 
of $\Delta\chi^2 \equiv \chi^2 - \chi_{\rm min}^2$ curves) for the five 
double-Gaussian parameters.  The best-fit value of each parameter is 
marked as well as the 90\% confidence limits.  
The $\Delta\chi^2$ curves have a deep minimum for all five parameters, an 
indication that the double-Gaussian model is a good description of 
the observed radial velocity distribution of inner spheroid 
stars.

\begin{deluxetable*}{lrrrrr}
\tablecolumns{6}
\tablewidth{0pc}
\tablecaption{Radial Velocity Distributions: Best Fit Gaussian Parameters.}
\tablehead{\multicolumn{1}{c}{Field}           & \multicolumn{5}{c}{Best fit Gaussian Parameters\tablenotemark{a}} \\
 & \multicolumn{2}{c}{Cold Component} & \multicolumn{2}{c}{Hot Spheroid~\tablenotemark{b}}
 & \multicolumn{1}{c}{Fraction}\\
 & \multicolumn{1}{c}{\mvsb} & \multicolumn{1}{c}{\sigvsb} 
 & \multicolumn{1}{c}{\mvsph} & \multicolumn{1}{c}{\sigvsph} 
 & \multicolumn{1}{c}{$N_{\rm sub}/N_{\rm tot}$} }
\startdata
All fields & $-285.4^{+12.8}_{-12.4}$ & $42.2^{+12.5}_{-14.3}$ & $-287.2^{+8.0}_{-7.7}$ & $128.9^{+7.7}_{-6.9}$ & $0.19^{+0.09}_{-0.08}$ \\
f109 & ... & ... & $-274.5^{+15.4}_{-15.3}$ & $120.7^{+11.7}_{-10.1}$ & ...\\ 
H11 & $-294.3^{+17.3}_{-17.6}$ & $55.5^{+15.6}_{-12.7}$ & $-287.2$ & 128.9 & $0.44^{+0.16}_{-0.16}$\\ 
f116 & $-309.4^{+19.2}_{-17.5}$ & $51.2^{+24.4}_{-15.0}$ & $-287.2$ & 128.9 & $0.44^{+0.22}_{-0.17}$ \\ 
f115 & ... & ... & $-270.9^{+18.6}_{-18.6}$ & $120.1^{+14.4}_{-12.0}$ & ... \\ 
f123 & $-279.4^{+5.1}_{-4.6}$ & \tablenotemark{c}~$10.6^{+6.9}_{-5.0}$ & $-287.2$ & 128.9 & $0.31^{+0.11}_{-0.11}$ \\ 
f135 & ... & ... & $-315.1^{+21.3}_{-21.3}$ & $127.8^{+16.5}_{-13.6}$ & ...\\ 
f130 & ... & ...  & $-259.8^{+19.3}_{-19.2}$ & $131.5^{+14.8}_{-12.5}$ & ...\\ 
a0 & ... & ... & $-299.2^{+25.2}_{-25.2}$ & $131.5^{+29.5}_{-21.6}$ & ...\\ 
\enddata
\tablenotetext{a}{A double Gaussian fit is presented for the combined 
sample, as it is a poor fit to a single Gaussian (\S\,\ref{sec:anal_comb}).  
Constrained double Gaussian fits are presented for three of the fields 
(H11, f116, and f123), with the wider component held fixed (adopting 
the fit to the combined sample).  Single Gaussian fits are presented 
for the remaining five fields.  The reader is referred to 
\S\,\ref{sec:anal_ind} for details of the fits to individual fields.
Errors quoted 
represent the 90\% confidence limits from the maximum-likelihood analysis.}
\tablenotetext{b}{The M31 RGB sample used in this analysis was chosen to ensure a high degree of completeness, and thus suffers from some MW dwarf contamination (\S\,A.2).  The MW dwarf star contaminants are largely at $v_{\rm hel}>-150$~\kms, and cause the best-fit \mvsph\ values to be biased towards more positive velocities.  The true \mvsph\ values of the M31 RGB population are 15 to 20~\kms\ more negative than listed here.  The effect on \sigvsph\ is negligible (\S\,A.3).}
\tablenotetext{c}{The median velocity error for the stars in f123 is 4.6~\kms.  The estimated intrinsic velocity dispersion of the cold component in f123 after accounting for velocity measurement error is 9.5~\kms.} 
\label{table:rvfits}
\end{deluxetable*}


\subsection{Maximum-Likelihood Fits to the Velocity Distributions of Individual Fields}\label{sec:anal_ind}

\begin{figure}
\plotone{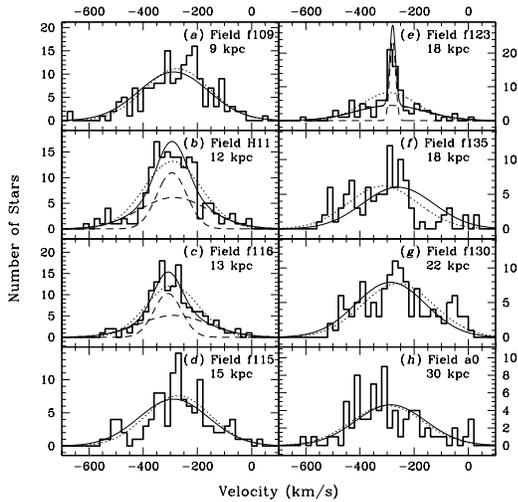}
\caption{
Velocity histograms for each of the individual fields with best-fit 
Gaussians overlaid.  Fields that did not show clear evidence of substructure
(f109, f115, f130, f135 and a0, \S\,\ref{sec:anal_ind}) are shown with 
the Gaussian component 
from the double-Gaussian fit to the 
full sample (\Gsph, \textit{solid curves}), as well as the best-fit single Gaussian to
the individual field (\textit{dotted curves}). 
Fields with evidence of substructure (H11, f116, 
f123), are shown with both their best-fit single (\textit{dotted curves}) 
and double (\textit{solid curves}) Gaussians.  For the constrained double-Gaussian 
fits, both the narrow and broad (\Gsph) Gaussian components are 
shown (\textit{dashed curves}) scaled to their relative contributions. 
}
\label{fig:velhist}
\end{figure}

As discussed in the previous section, the combined data set shows definite
evidence of a kinematically cold component centered at about $-300$~\kms\
that comprises a significant fraction (19\%) of the total population.  We next
investigate which of the fields in our data set are the main contributors to
this cold population.  Figure~\ref{fig:velhist} shows
velocity histograms for each of the eight fields analyzed in this paper.
While most of the fields display hints of substructure--- in the form of one
or more small possible peaks in their velocity distribution that may be
marginally significant relative to the (substantial) Poisson noise--- we are
specifically interested in judging each field's contribution to the cold
component at $\sim-300$~\kms.  For this purpose, we compare the data in each
field to three sets of Gaussian fits.  We describe each of the fits below,
summarize the results of the fits, and then list the quantitative details of
the fits for each field.

First, the radial velocity distribution in each field is compared to the
Gaussian \Gsph\ defined by \mvsph$=-287.2$~\kms,
\sigvsph$=128.9$~\kms\ (\S\,~\ref{sec:anal_comb}) using the reduced $\chi^2$ 
statistic ($\chi^2_{\rm red}$).  Fields f109, f115, and a0 are consistent with being drawn from 
\Gsph, and so are ruled out as significant
contributors to the $\sim-300$~\kms\ cold component.  
The rest of the fields are at least marginally inconsistent with being drawn
from \Gsph.

Second, a maximum-likelihood single-Gaussian 
fit is performed on the radial velocity distribution in each field and 
compared to the data.
Fields f123 and f135 are inconsistent with their respective best-fit single
Gaussians (based on the $\chi^2_{\rm red}$ statistic) and are therefore
suspected to contain substructure.  The remaining fields (f109, H11, f116,
f115, f130, and a0) are consistent with their best-fit single Gaussians.  For
fields f109, f115, f130, and a0, the best-fit Gaussians are consistent with
\Gsph, and they are thus ruled out as significant contributors to the
$\sim-300$~\kms\ cold component.  Fields f116 and H11 are suspected to
contain substructure because
their best-fit Gaussians are significantly narrower than \Gsph.  Previous
kinematic studies of M31's inner spheroid, including this one, have found a
ubiquitously broad distribution of radial velocities ($v_{\rm hel}\approx0$
to $\approx-600$~\kms; \S\,~\ref{sec:veldisp}). 
Thus, the anomalously narrow single Gaussian fits in
fields f116 and H11 cause us to suspect them of being contributors to the
$\sim-300$~\kms\ cold component.

Third, we carry out a maximum-likelihood fit to all fields using a
constrained double Gaussian, with \Gsph\ as the fixed wide Gaussian component.
Fields H11, f116, and f123 are well fit by a constrained double Gaussian
(based on the $\chi^2_{\rm red}$ statistic and well-defined minima for the
variable Gaussian parameters).  These three fields are significant
contributors to the $\sim-300$~\kms\ cold component, and the constrained
double-Gaussian fit is adopted as the preferred fit in the subsequent
discussion.  The best-fit cold component in the constrained double-Gaussian
fit to f130 is centered at $\sim-50$~\kms, but this likely represents
residual contamination by MW dwarf stars (\S\,\ref{sec:sample},
\S\,A).  The remaining fields (f109, f115,
f135, and a0) are poor fits to a constrained double-Gaussian
in that the Gaussian parameters do not have well-defined minima.

In summary:
\begin{itemize}
\item{Three fields, H11, f116, and f123 are identified as significant
contributors to the $\sim-300$~\kms\ cold component.}
\item{Although field f135 shows evidence of substructure, it is a poor fit
to a constrained double-Gaussian and is not a definite contributor to
the $\sim-300$~\kms\ cold component.  (However, the fit may be confused by
the presence of multiple cold components; \S\,\ref{sec:f135}.)}
\item{Four fields, f109, f115, f130, and a0, are not significant
contributors to the $\sim-300$~\kms\ cold component.}
\end{itemize}

Each field is discussed individually below, and Table~\ref{table:rvfits} 
summarizes the preferred fits (single or double-Gaussian) to the velocity
distributions in each field.   

\noindent\textit{Field f109}: The data in this field are consistent with being drawn
from the Gaussian \Gsph\
(\S\,\ref{sec:anal_comb}).  The best-fit single Gaussian to the data in 
this field has parameters \mv$=-274.5$~\kms\ and $\sigma=120.7$~\kms, and the data
are also consistent with being drawn from this distribution.

\noindent\textit{Field H11}: The probability the data in this field are drawn
from the Gaussian \Gsph\
is $P<1$\%, thus the data are inconsistent with this distribution. 
The best-fit single Gaussian to the data in 
this field has parameters \mv$=-291.1\pm 11.6$~\kms\ and $\sigma=106.2^{+8.7}_{-7.7}$~\kms, and the data
are consistent with being drawn from this distribution.  However, the best-fit value
of $\sigma$ in this field is anomalously low compared to the value of
\sigvsph\ determined from the double-Gaussian fit to the combined data set: 
the two values are inconsistent at the $\sim 3.5\sigma$ level.  This
suggests that there is a kinematically cold population in this field,
and a comparison of the data to the constrained double-Gaussian fit to 
this field 
(\mvsb$=-294.3$~\kms, \sigvsb$=55.5$~\kms, $N_{\rm sub}/N_{\rm tot}=0.44$) 
returns a $\chi^2_{\rm red}$ 
that is significantly smaller than the $\chi^2_{\rm red}$ of the single
Gaussian fit.
 
\noindent\textit{Field f116}: The data in this field are inconsistent
with being drawn from the Gaussian \Gsph.  The best-fit single 
Gaussian to the data in 
this field has parameters \mv$=-292.9^{+11.5}_{-11.6}$~\kms\ and $\sigma=97^{+8.7}_{-7.5}$~\kms, and the data
are consistent with being drawn from this distribution.  As in field H11, the
best-fit value of $\sigma$ is inconsistent with \sigvsph\ from the combined
data set, at the $\sim 4.2\sigma$ level.  A comparison of the data to the 
constrained double-Gaussian 
fit to this field (\mvsb$=-309.4$~\kms, \sigvsb$=51.2$~\kms,  
$N_{\rm sub}/N_{\rm tot}=0.44$) returns a significantly smaller $\chi^2_{\rm red}$ 
than that of the single-Gaussian fit.
 
\noindent\textit{Field f115}: The data in this field are consistent with being drawn
from the Gaussian \Gsph, as well as the best-fit single Gaussian
(\mvsph$=-270.9$~\kms, \sigvsph$=120.1$~\kms).

\noindent\textit{Field f123}: The probability that the data in this field are
drawn from the Gaussian \Gsph\ is $P<\!\!< 1$\%.
The probability that the data are drawn 
from the best-fit single Gaussian (\mvsph$=-270.9$~\kms, \sigvsph$=120.1$~\kms) in
this field is also $<\!\!< 1$\%.  The data in this field are strongly inconsistent with being drawn from any single Gaussian, but they
are consistent with the constrained double-Gaussian fit (\mvsb$=-279.4$~\kms, \sigvsb$=10.6$~\kms, $N_{\rm sub}/N_{\rm tot}=0.31$).  The median velocity error for the stars in field f123 is 4.6~\kms, therefore the intrinsic velocity dispersion of the cold component in this field is estimated to be 9.5~\kms.

\noindent\textit{Field f135}: The data in this field are inconsistent with being drawn
from the Gaussian \Gsph\ ($P\sim 1$\%). The best-fit single Gaussian to the data in 
this field has parameters \mvsph$=-315.1$~\kms\ and \sigvsph$=127.8$~\kms, but the data
are also inconsistent with being drawn from this distribution ($P\lesssim1$\%).
However, the maximum-likelihood analysis was unable to constrain a double-Gaussian 
fit to any reasonable degree of certainty--- i.e., the error estimates on the 
parameters show no strong global minima.  Field f135 is therefore treated 
as a field without a {\it definite\/} detection of substructure, although in 
\S\,\ref{sec:f135} we discuss the possible presence of multiple 
kinematically-cold components in this field.  

\noindent\textit{Field f130}: The data in this field are inconsistent with being drawn
from the Gaussian \Gsph\ ($P\lesssim 1$\%), but are consistent with being drawn
from the best-fit single Gaussian (\mvsph$=-259.8$~\kms,
\sigvsph$=131.5$~\kms).  The best-fit single Gaussian for this field has 
parameters which are consistent with \Gsph\ (Table~\ref{table:rvfits}).

\noindent\textit{Field a0}: The data in this field are consistent with 
being drawn from the Gaussian \Gsph\, as well as the
best-fit single Gaussian (\mvsph$=-299.2$~\kms, \sigvsph$=131.5$~\kms).

\bigskip
We now have a characterization of the kinematical properties of the
combined data set and the individual fields, and have identified the fields
which significantly contribute to the $\sim-300$~\kms\ cold component discovered
in the combined data set (H11, f116, and f123).  The number of stars in fields
H11, f116, and f123 which are associated with the cold component comprise
17\% of the total number of M31 RGB stars in the sample, confirming that
these fields are the primary contributors to the cold component.
Next, we discuss the trends in the properties of
the dynamically hot and cold populations (\S\,\ref{sec:veldisp}
and \S\,\ref{sec:coldpop}, respectively), followed by possible
physical interpretations of
the cold component (\S\,\ref{sec:origin}).

\section{Velocity Dispersion of M31's Virialized Inner Spheroid}\label{sec:veldisp}
As discussed in \S\,\ref{sec:kin}, the radial velocity distribution of the
combined data set contains a significant cold component at $v_{\rm hel}\sim -300$~\kms.  An analysis of the 
kinematical profile of the individual fields shows that this is due to the
presence of a significant amount of substructure in three of the fields (\S\,\ref{sec:anal_ind}).
The kinematically hot component in the double
Gaussian fit to the combined data set presumably represents the 
underlying virialized inner spheroid of M31.  
The velocity dispersion of the spheroid based on a maximum-likelihood double 
Gaussian fit to the 
combined data set (1013 M31 RGB stars) is \sigvsph = 128.9~\kms.

We have combined all of the fields to get a more robust estimate of the 
velocity dispersion of the inner spheroid, since the individual fields suffer
from small number statistics.  However, the dynamical quantities 
(\mvsph, \sigvsph) may have radial dependencies.  The velocity
dispersion of spheroids is expected to decrease with increasing radius 
\citep[e.g.,][]{nav96,nav04,mam05,dek05}.  Rotation of the spheroid or 
the tangential motion of M31 could result in a radial dependency of \mvsph, 
but since our fields are mostly aligned along the minor axis and cover a 
relatively small radial range, we do not expect to be sensitive to these
effects.

\begin{figure}
\plotone{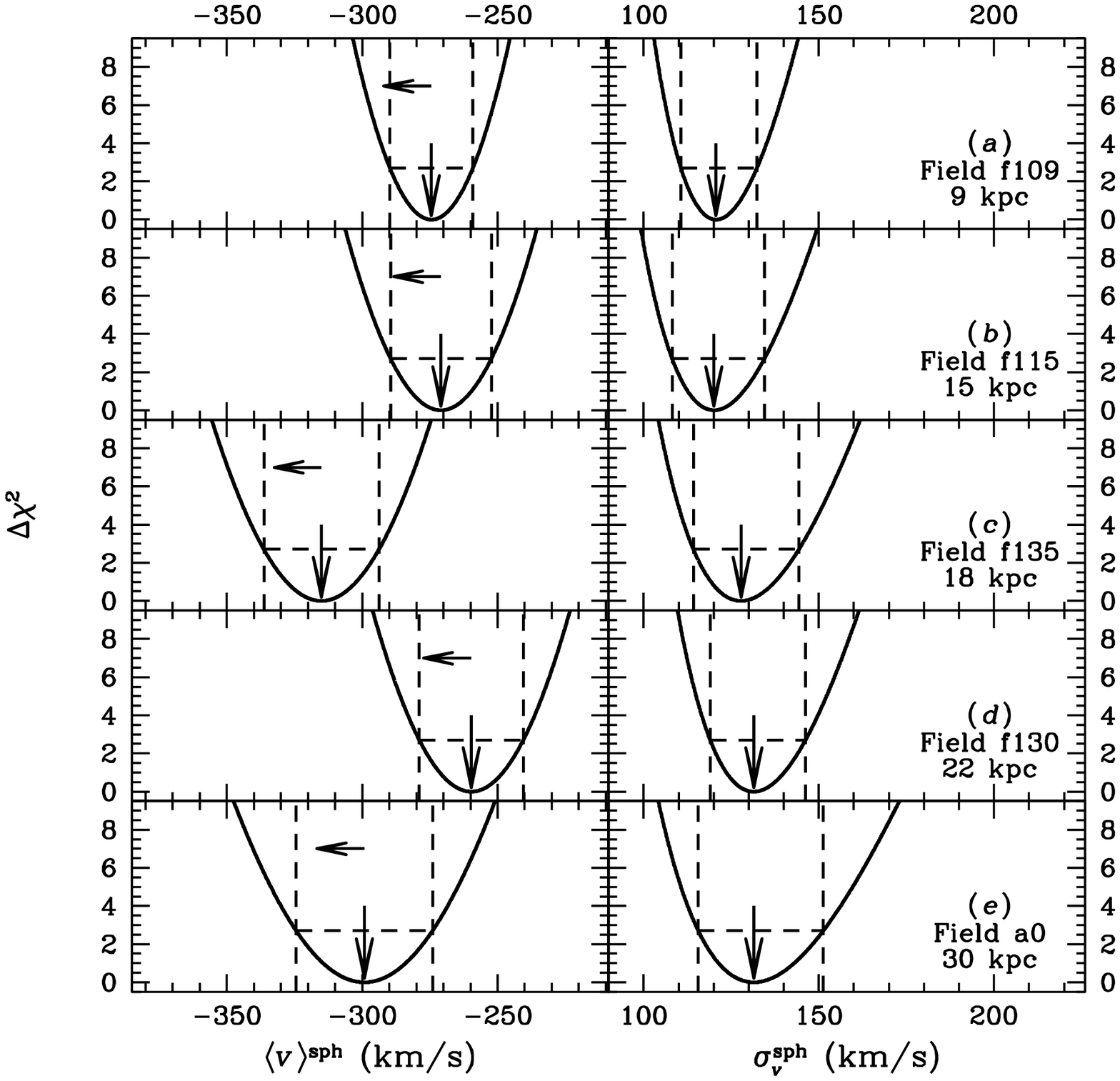}
\caption{
Results of the maximum-likelihood analysis for the
single-Gaussian fits to individual 
fields that do not show clear evidence of the $\sim -300$~\kms\ cold component in 
their radial velocity distributions (\S\,\ref{sec:anal_ind}). 
The panels show $\Delta\chi^2$ for the mean velocity \mvsph\ (\textit{left}) and 
velocity dispersion \sigvsph\ 
(\textit{right}) of M31 RGB stars in fields (\textit{a}) f109, (\textit{b}) 
f115, (\textit{c}) f135, (\textit{d}) f130, and (\textit{e}) a0.  As in 
Figure~\ref{fig:errors_all}, the optimal values of \mvsph\ 
and \sigvsph\ are marked by arrows, and the 90\% confidence limits are shown
as dashed lines. 
The horizontal arrow in each of the left panels
represents the correction (magnitude and direction) to the \mvsph\ value necessary to offset the
bias caused by MW dwarf contamination at $v_{\rm hel}>-150$~\kms\ (\S\,A.3). 
}
\label{fig:errors_sg}
\end{figure}

To test for a dependency of the dynamical quantities on radius, we analyze the single
Gaussian fits to the velocity distributions in each of the five fields which 
do not show 
clear evidence of substructure as described in \S\,\ref{sec:anal_ind} 
(Table~\ref{table:rvfits}).  The $\Delta\chi^2$ 
error estimates for the best-fit single Gaussian parameters in these fields are shown in 
Figure~\ref{fig:errors_sg}.
The best-fit \mvsph\ values in each field are largely consistent with
the systemic velocity of M31 ($v_{\rm sys}=-300$~\kms) once the effect of
MW dwarf star contamination is taken into account,
with the exception of f135, whose 
\mvsph\ is significantly more negative.
Our data show no evidence for a decreasing \sigvsph\ with increasing radius. 
However, the fields discussed in this paper only span a range 
in projected radial distance of $R_{\rm proj}\sim9$ to 30~kpc.  This is small 
compared to the size of the total M31 spheroid (inner spheroid and outer halo), which has been shown
to extend to 165 kpc \citep{gil06}.  
\citet{bat05} use a sample of 240 Galactic halo objects to determine the 
radial velocity dispersion of the Milky Way halo, and find that it has an
almost constant value of 120~\kms\ out to 30 kpc, beyond which it decreases
with increasing radial distance, declining to 50~\kms\ at 120 kpc. 

There have been a few previous
measurements of the velocity dispersion based on M31 spheroid stars. 
The closest analog to the present study is \citet{rei02}, who fit 80 
candidate M31 RGB stars at $R_{\rm proj}=19$ kpc on the southeastern minor axis 
with a combination of the 
Galactic standard model and a wide Gaussian. They found \sigvsph
$\sim 150$~\kms\ for the M31 velocity dispersion, with the number of 
M31 RGB stars estimated to be 43\% of the population.  
\citet{guh06} studied a field on the giant
southern stream at $R_{\rm proj}=33$ kpc and found a velocity dispersion
of \sigvsph$=65^{+32}_{-21}$~\kms\ for the underlying spheroid
based on a sample of $\approx21$ 
stars.  However, if 3 likely RGB stars with $v_{\rm hel}>-150$~\kms\ are included,
the estimated velocity dispersion in this field increases to \sigvsph$=116^{+31}_{-22}$~\kms.

\citet{cha06} have measured a mean velocity dispersion of \sigvsph$=126$~\kms\ 
for the inner spheroid
of M31 using $\sim 800$ RGB stars in multiple fields 
surrounding M31. They determine that
the spheroid has a central velocity dispersion of 152~\kms\ which decreases by 
$-0.9$~\kms~kpc$^{-1}$ out to $R_{\rm proj}\sim 70$~kpc.  Many of their fields are near
M31's major axis and have significant contamination from the extended 
rotating component identified as the extended disk in \citet{iba05}.  
They isolate a sample of 
M31 spheroid stars by removing all stars within 160~\kms\ of the disk 
velocity in each field, and by 
removing all stars with $v_{\rm hel} > -160$~\kms\ (MW dwarf star contaminants).  
This ``windowing'' technique leaves them with a sample of M31 spheroid stars that is 
significantly incomplete, but is largely uncontaminated by M31's extended disk or MW dwarf stars.  
Based on the \citet{cha06} result, we would expect to measure a velocity 
dispersion of 146~\kms\ in our innermost field, f109, which would decrease to
a velocity dispersion of 125~\kms\ in our outermost field, a0.  The predicted
velocity dispersion of 146~\kms\ in field f109 exceeds the 90\% confidence
limits of the maximum-likelihood fit and is just within the 99\% confidence limit, 
and the data presented in this paper show no evidence of a strong trend 
in \sigvsph\ with radius.

The previous measurements of the stellar velocity 
dispersion of the M31 spheroid
relied either on samples that were chosen on the basis of radial velocity cuts 
or on a statistical 
fit to the combined M31 RGB and MW dwarf populations.  Our measurement
of the velocity dispersion
of M31's spheroid is unique in that it is based on a sample of spectroscopically confirmed 
M31 RGB stars that were chosen
\textit{without} the use of radial velocity (\S\,\ref{sec:sample}).   
Our method also allows us
to quantify and correct for the effect of MW dwarf star contamination (\S\,A.2 and \S\,A.3).

\section{Properties of the Minor-Axis Substructure}\label{sec:coldpop}

\subsection{Spatial Trends: Kinematics and Structure}\label{sec:spatdist}
 
\begin{figure}
\plotone{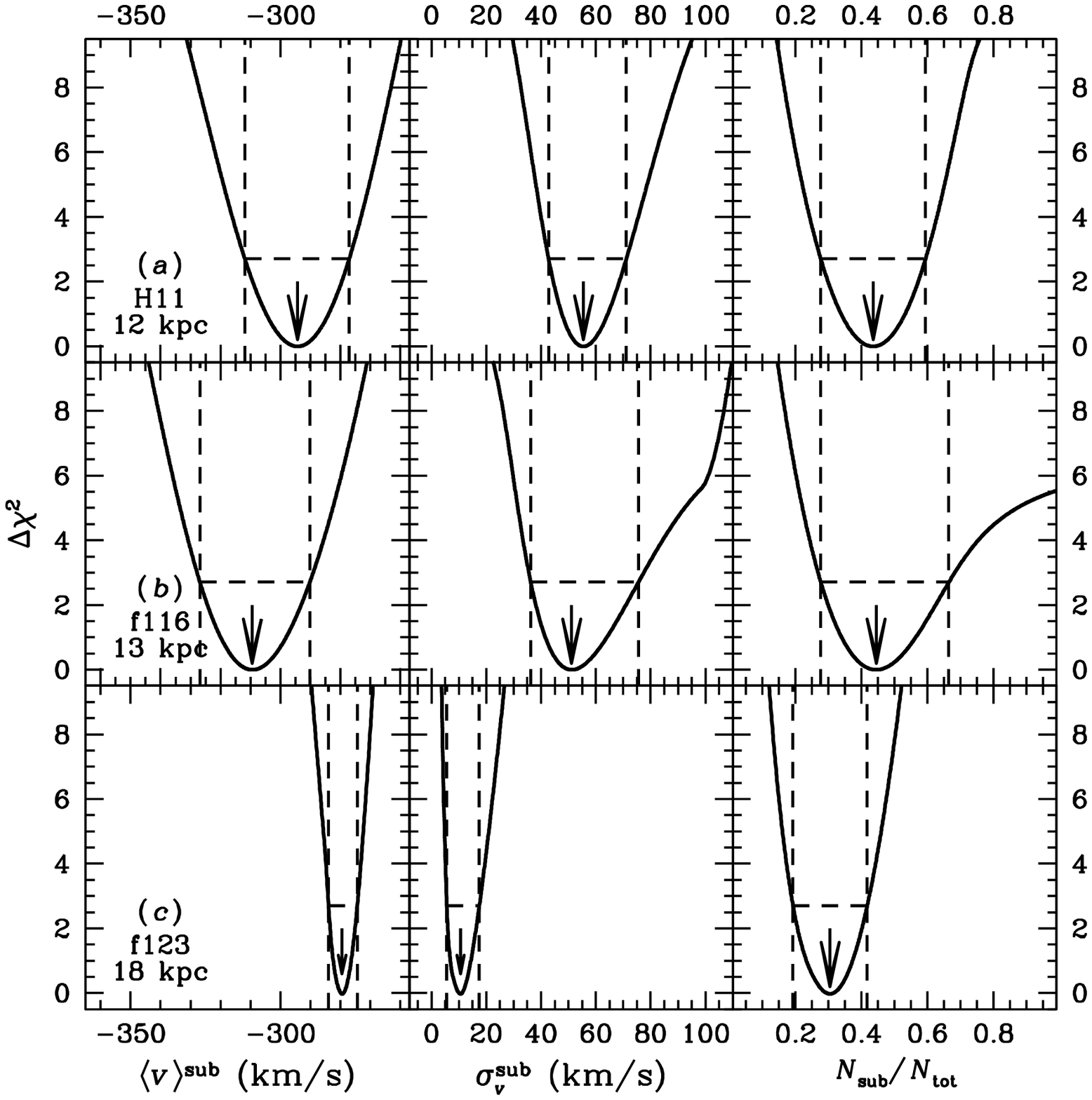}
\caption{
Results of the maximum-likelihood analysis for the
narrow Gaussian parameters 
from the constrained double-Gaussian fits to fields (\textit{a}) H11, 
(\textit{b}) f116, and (\textit{c}) f123.
The panels show $\Delta\chi^2$ as a function of (\textit{left}) 
mean velocity \mvsb, (\textit{middle}) velocity dispersion \sigvsb,
and (\textit{right}) fraction of stars in the cold component, $N_{\rm sub}/N_{\rm tot}$. 
The parameters of the wide Gaussian component in these fits have been fixed at 
the values of the Gaussian \Gsph\ (\S\,\ref{sec:anal_comb}).  
As in Figure~\ref{fig:errors_all}, the optimal values of each parameter
are marked by arrows, and the 90\% confidence limits are shown
as dashed lines.  The velocity dispersion decreases with increasing radial 
distance from the center of M31.
}
\label{fig:errors_sub}
\end{figure}

As discussed in \S\,\ref{sec:anal_ind}, fields H11, f116, and f123 show 
evidence of the cold component at $\sim-300$~\kms\ in their radial velocity
distributions 
(Fig.~\ref{fig:velhist}), and are well fit by a sum of two Gaussians with 
\Gsph\ as the fixed wide Gaussian component (Table~\ref{table:rvfits}).  The 
$\Delta\chi^2$ error estimates for the  free parameters 
(\mvsb, \sigvsb, and $N_{\rm sub}/N_{\rm tot}$) are shown in 
Figure~\ref{fig:errors_sub} for each field.  

If the three fields are considered together, a pattern emerges.  Both the 
velocity dispersion and the fraction of stars in the cold component 
decrease with increasing radial distance.  The cold component in fields H11
and f116 (at $R_{\rm proj}=12$ and 13 kpc, respectively) is significantly 
wider than the cold component in field f123 ($R_{\rm proj}=18$ kpc).
The cold component also appears to be more dominant by number (surface
brightness) over the hot component in fields H11 and f116 
than in field f123.   

\begin{figure}
\plotone{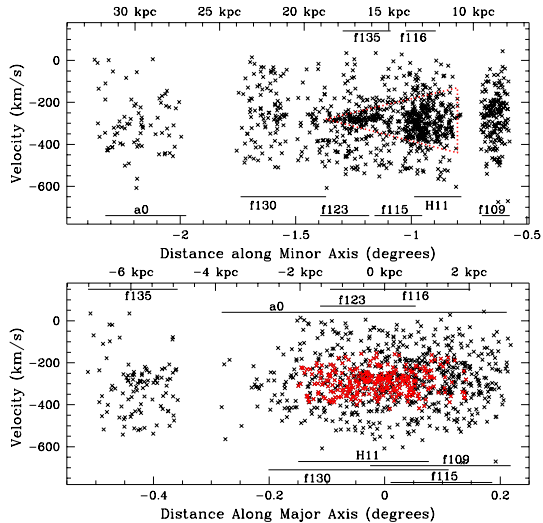}
\caption{
Distribution of M31 RGB stars in velocity vs. distance along the minor 
(\textit{top}) and major (\textit{bottom}) axes
of M31.  The range of minor (major) axis distances
of stars in each field are shown in the top (bottom) panel.  \textit{Top:}  The 
cold component is visible as a triangular feature that starts 
at $-0.8$\degree\ (10.9~kpc) and narrows to a point at $-1.35$\degree\ (18.4~kpc) along the minor axis.  
This feature is outlined in red (\textit{dotted line}).  \textit{Bottom:}
  Stars in fields H11, f116, and f123 that fall within the 
triangular outline in the top panel are colored red.  In the fields in
which the $\sim-300$~\kms\ cold component is present, it is spread evenly 
as a function of projected distance from the minor axis.
}
\label{fig:spatdist}
\end{figure}

Figure~\ref{fig:spatdist} presents the velocities of the M31 RGB stars as a
function of their distance along the major and minor axes of M31.  The cold
component can be seen as a triangular-shaped feature that narrows to
a sharp point as the distance along the minor axis increases. The fields that 
overlap the triangular-shaped feature in minor axis distance 
are (in order of increasing distance along the minor axis) H11, f116, f115, 
f135, and f123.   Stars in fields
H11, f116 and f123 that are within the 
area denoted by the dotted line are shown as red crosses in the bottom panel. 
In the fields
in which it is observed, the cold component is spread evenly along the 
direction of the major axis, and is centered at \vhel$ \sim -285$~\kms.
The majority of the fields overlap in position along the major axis; the 
exception is field f135, which is the isolated set of points at 
large major axis distances
($-0.51$\degree\ to $-0.36$\degree, or $-7.0$~kpc to $-4.9$~kpc) in the bottom panel of Figure~\ref{fig:spatdist}.   
Field f123 extends from $-1.18$\degree\ 
to $-1.39$\degree\ (16.1~kpc to 19.0~kpc) along the minor axis; the tip of the feature is seen in this field.  
This field also has the coldest substructure in the radial velocity 
histograms (Fig.~\ref{fig:velhist}, Table~\ref{table:rvfits}).  Field H11 
brackets the other edge of the feature along the minor axis.  Field H11 has a range in minor 
axis distance of $-0.78$\degree\ to $-0.99$\degree\ (10.7~kpc to 13.5~kpc), and has a large degree of 
overlap with field f116 in both 
their minor and major axis distance ranges.  The fits to the cold component in
these two fields return similar \sigvsb\ and $N_{\rm sub}/N_{\rm tot}$ estimates
(Table~\ref{table:rvfits}).

Of the fields in which substructure was not clearly detected kinematically
(\S\,\ref{sec:anal_ind}), fields f135 and f115 are both within the minor axis
distance range spanned by the observed substructure (dotted triangular region
in upper panel of Fig.~\ref{fig:spatdist}).
A two-sided Kolmogorov-Smirnov (KS) test finds that the radial velocity 
distribution of field f115 is consistent
with the radial velocity distribution of its closest neighbor, field f116 
(Fig.~\ref{fig:cfht}), although it is inconsistent with the best
constrained double Gaussian fit to field f116.  Although the radial velocity distribution of 
field f135 is not well-fit by a double Gaussian (\S\,\ref{sec:anal_ind}), 
there is a concentration of stars near $v_{\rm hel}\sim -300$~\kms\ 
in its radial velocity distribution (Fig.~\ref{fig:velhist}).  We will
discuss these fields further in the context of the physical interpretation
of the cold substructure (\S\,\ref{sec:gss}).  Fields f130 and a0 are at 
larger minor axis distances than the tip of the 
feature, and field f109 is interior to the feature.  The fact that 
substructure is not detected in these fields is consistent with our favored
physical interpretation of the $-300$~\kms\ cold component, which will be discussed in \S\,\ref{sec:gss}.

\subsection{Metallicity Distribution}\label{sec:met}
So far we have considered only the kinematic properties of the minor axis
population.  The distribution of stellar metallicities, however, is also
a powerful diagnostic of the presence and origin of substructure, since 
different galactic components (e.g., disk, inner spheroid) and tidal debris
have different formation histories, and therefore different chemical
abundances.  

\begin{figure}
\plotone{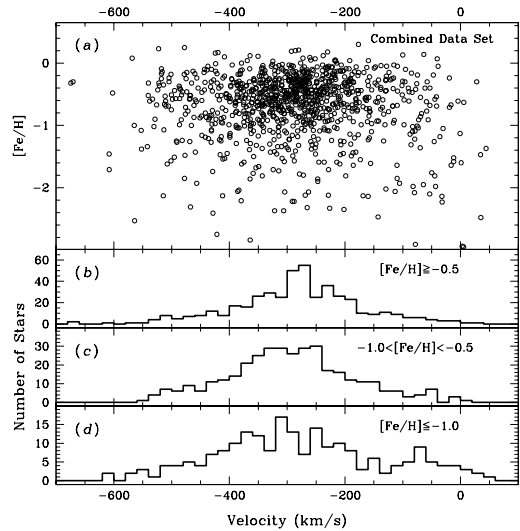}
\caption{
(\textit{a}) Metallicity vs. heliocentric velocity for the full M31 RGB sample.
The majority of the population is metal-rich, with an evenly distributed 
metal-poor tail over the full range of velocities.  A concentration of 
metal-rich stars 
near $v_{\rm hel}=-300$~\kms\ can be seen. 
The bottom three panels show velocity histograms for subsets of the data: 
(\textit{b}) \fehp$\ge -0.5$, (\textit{b}) $-1.0<$\fehp$< -0.5$, and 
(\textit{d}) \fehp$\le -1.0$.  The $\sim-300$~\kms\ cold component is more dominant in 
the metal-rich samples.  A maximum-likelihood double-Gaussian fit was 
performed for each subset
of the data, with all of the parameters except $N_{\rm sub}/N_{\rm tot}$ 
held fixed at the best-fit values for the complete M31 RGB sample 
(\S\,\ref{sec:anal_comb}, Fig.~\ref{fig:vel_all}, Table~\ref{table:rvfits}).  
The cold component
comprises a negligible fraction of the population in (\textit{d}), 20.9\% of the population
in (\textit{c}), and 29.2\% of the population in (\textit{b}), indicating that
the $\sim-300$~\kms\ cold component is relatively metal-rich.  
}
\label{fig:vel_met}
\end{figure}

Figure~\ref{fig:vel_met}\,(\textit{a}) displays \fehp\ vs. $v_{\rm hel}$ 
for the M31 RGB stars.  The \fehp\ values
are based on a comparison of the star's position within the \ivi\ 
CMD to a finely spaced grid of theoretical 12 Gyr, [$\alpha$/Fe]~$=0$ 
stellar isochrones \citep{kal06b,vdb06} adjusted to the distance of M31 
\citep[783~kpc;][]{sta98,hol98}.  
Stars with \fehp$ <-1$ appear to be
evenly distributed in velocity, while there is an obvious clump of 
metal-rich stars with velocities near $-300$~\kms.  The bottom panels 
(\textit{b--d}) show velocity histograms for stars in three \fehp\ bins: 
(\textit{b}) \fehp$\ge-0.5$, (\textit{c}) $-1.0<$\fehp$<-0.5$, and 
(\textit{d}) \fehp$\le-1.0$.  The strength of the cold component in each 
metallicity bin is measured by performing a maximum-likelihood double-Gaussian
fit to the velocity distribution.  Only the fraction of stars in the 
cold component ($N_{\rm sub}/N_{\rm tot}$) is allowed to vary; the rest
of the parameters are held fixed at the best-fit values from the fit to the
full M31 RGB sample (\S\,\ref{sec:anal_comb}, Table~\ref{table:rvfits}).  The fraction of stars 
in the cold component is 29.2\% in (\textit{b}), 20.9\% in (\textit{c}), and
is negligible in (\textit{d}), indicating that the cold component is
metal-rich relative to the dynamically hot component.     

\begin{figure}
\epsscale{0.9}
\plotone{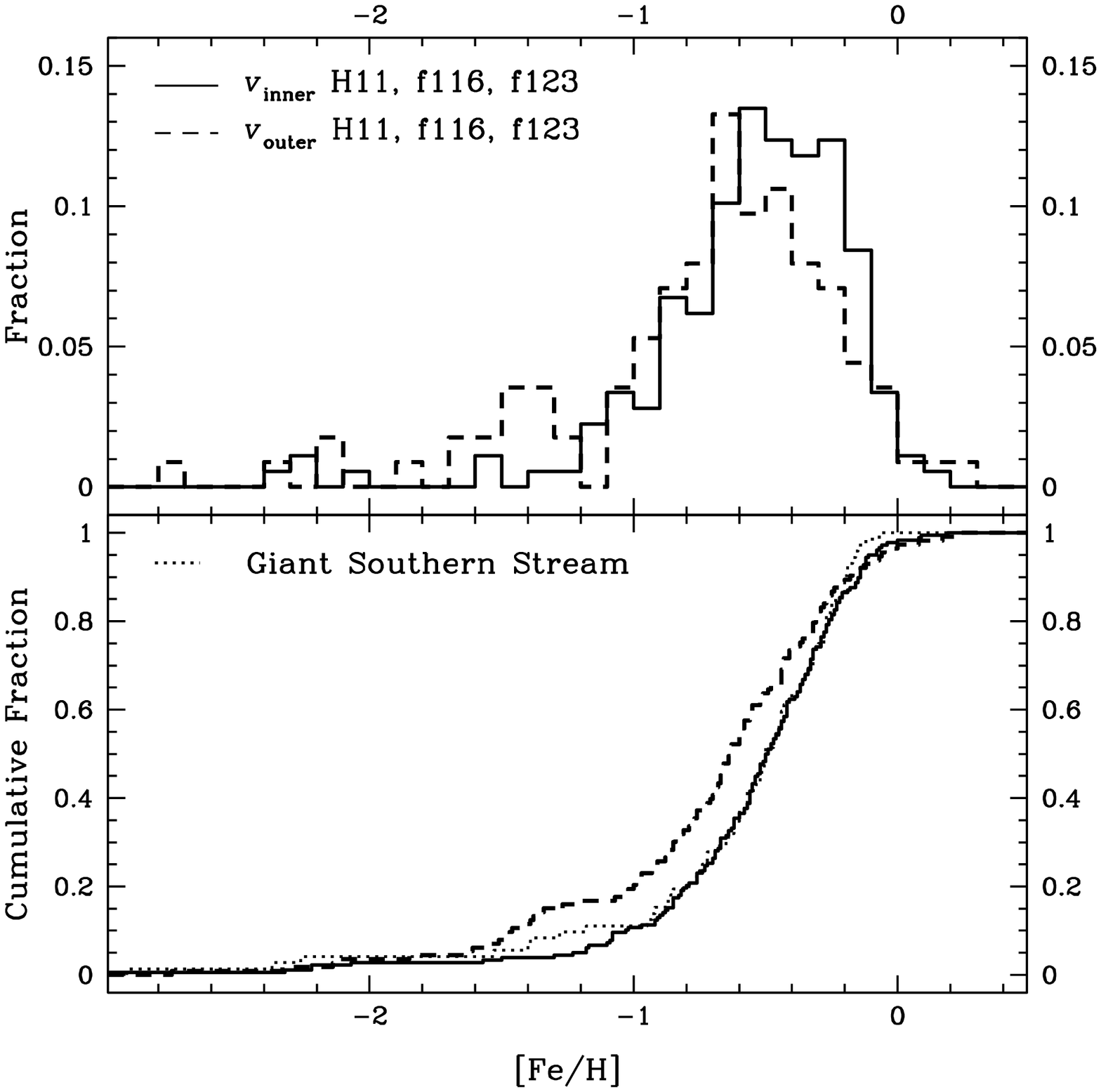}
\caption{
The \fehp\ distribution of the $v_{\rm inner}$ (\textit{solid line}) and $v_{\rm outer}$ 
(\textit{dashed line}) components 
in fields H11, f116, and f123, in histogram (\textit{top}) and cumulative 
(\textit{bottom}) form.  The $v_{\rm inner}$ (cold component)
sample is defined to be stars with velocities within the range 
\mvsb~$-$~\sigvsb~$<v_{\rm hel}<$~\mvsb~$+$~\sigvsb, where \mvsb\ and \sigvsb\ represent the best-fit
narrow Gaussian components in each field.   This range was chosen to maximize the
percentage of substructure stars compared to hot spheroid stars.  
The $v_{\rm outer}$ (hot spheroid) sample consists of stars with 
velocities $v_{\rm hel}<$~\mvsb~$- 2$\sigvsb\ or ~\mvsb~$+ 2$\sigvsb$<v_{\rm hel}<-150$~\kms.  
This minimizes contamination from substructure stars and MW dwarf stars (\S\,~A.2) in the hot spheroidal \fehp\ 
distribution.  
The $v_{\rm inner}$ sample is slightly more metal rich than the $v_{\rm outer}$ sample. 
  Since the $v_{\rm inner}$ sample is contaminated
by spheroid stars, the true difference in \fehp\ between the substructure and spheroid
populations is somewhat greater than indicated in this plot.
The cumulative \fehp\ distribution of stars in M31's
GSS \citep[from fields at 33 and 21~kpc;][]{guh06,kal06a} 
is plotted in the bottom panel for comparison (\textit{thin dotted line}), 
and will be discussed in \S\,\ref{sec:gss}.  
}
\label{fig:met}
\end{figure}

Figure~\ref{fig:met} compares the metallicity distributions of stars in the velocity
range of the $\sim-300$~\kms\ cold component (\textit{solid line}) discovered in 
fields H11, f116, and f123 and stars that are identified with the hot spheroidal
component (\textit{dashed line}) in those fields.  The hot 
spheroidal distribution ($v_{\rm outer}$) is based on stars whose velocities are greater than
$\pm 2$\sigvsb\ away from \mvsb.  This minimizes contamination of the 
spheroidal component by stars associated with the $\sim-300$~\kms\ cold
component.  An additional
constraint on the $v_{\rm outer}$ distribution is that only stars with 
$v_{\rm hel}<-150$~\kms\ are included in order to avoid residual
MW dwarf star contaminants that lie in the range
$v_{\rm hel}>-150$~\kms\ (\S\,A.2).  
Stars within
the velocity range of the cold component can 
only statistically be identified as belonging to the hot spheroid or
cold component, thus it is not possible to identify an
uncontaminated sample of the $\sim-300$~\kms\ cold component.  Stars with
velocities within
$\pm1$\sigvsb\ of \mvsb\ are used for the $v_{\rm inner}$  
\fehp\ distribution, in 
order to maximize the number of members of the cold component while minimizing 
the contribution of spheroid stars.  

The \fehp\ distribution of the $v_{\rm outer}$ sample has a 
peak at lower \fehp\ values and a larger metal-poor tail than the \fehp\
distribution of the $v_{\rm inner}$ sample.  The mean (median) metallicity of the 
$v_{\rm outer}$ sample is $\langle$\fehp$\rangle_{\rm mean}=-0.72$
($\langle$\fehp$\rangle_{\rm med}=-0.63$), while the mean (median) metallicity of 
the $v_{\rm inner}$ sample is $\langle$\fehp$\rangle_{\rm mean}=-0.55$
($\langle$\fehp$\rangle_{\rm med}=-0.49$).  A KS test
returns a probability of 0.7\% that the two distributions are drawn from the 
same parent distribution.  The $v_{\rm inner}$ distribution
is highly contaminated by spheroid stars even within $\pm1$\sigvsb\ of \mvsb\ 
(Fig.~\ref{fig:velhist}); the estimated contamination of the $v_{\rm inner}$ sample
by spheroid stars is 32.5\%.  
A statistical subtraction of the $v_{\rm outer}$ distribution (scaled by
the contamination rate) from the 
$v_{\rm inner}$ distribution yields a distribution with a mean (median) metallicity of 
$\langle$\fehp$\rangle_{\rm mean}=-0.52$ 
($\langle$\fehp$\rangle_{\rm med}=-0.45$). 

An important effect on the measured \fehp\ values is the assumed age
(12 Gyr) of the stellar population.  This assumption is wrong for at least 
field H11, which has been shown by deep HST/ACS imaging to contain a 
wide spread of stellar ages, ranging from 6\,--\,13.5 Gyr 
\citep{bro03,bro06b}.  Our data 
show a sharp cutoff in the \fehp\ distribution at \fehp$\sim0$, with a 
metal-poor tail that extends out to \fehp$\sim-1.5$ to $-2.0$ 
(Fig.~\ref{fig:met}).  The \citet{bro06b} HST/ACS data show a metal-rich 
cutoff in the \fehp\ distribution at \fehp$\sim0.3$, with a metal-poor tail 
which extends to \fehp$\sim-1.5$ to $-2.0$ (Fig.~9 of \citet{bro06b}).  
The star formation history derived from the HST/ACS CMDs show that the 
intermediate age population (6\,--\,9~Gyr) is metal-rich (\fehp$\sim0$), 
while the old (10\,--\,14 Gyr) stellar population is relatively metal-poor. 
Thus, our assumed age of 12 Gyr will underestimate metallicities for stars with 
\fehp\ values near solar, but is appropriate for 
the more metal-poor stars (\fehp$\lesssim-0.5$).
Varying the age between 6 and 14~Gyr introduces an $\approx 0.3$--$0.4$ dex 
spread in the \fehp\ values derived from the isochrone fitting \citep{kal06b}; after 
accounting for this offset at the metal-rich end of the \fehp\ distribution, 
our \fehp\ distribution is consistent with the \citet{bro06b} result.


The intrinsic spread in ages in the spheroid found by \citet{bro03,bro06b} can
have two possible effects on our comparison of the \fehp\ distributions 
of the $v_{\rm inner}$ and $v_{\rm outer}$ samples. 
If all the stars in these fields have a common spread 
in ages regardless of their kinematical properties, the error in the \fehp\ 
measurement introduced by
assuming a uniform age for the population will cause a shift in the actual
\fehp\ values, but the relative difference between the $v_{\rm inner}$ and $v_{\rm outer}$
populations will not be greatly affected.  If the stars associated with the 
$\sim-300$~\kms\ cold component
are systematically younger than the underlying smooth spheroid 
population, 
the measurement of \fehp\ for the $v_{\rm inner}$ sample derived from the 12 Gyr isochrones will be biased 
towards low metallicities, and 
thus the intrinsic difference 
in metallicity between the two populations will be greater than shown.  

We have also
assumed that all stars are at the same line-of-sight distance as M31's center.  This is
a valid approximation for the inner spheroid: at $R_{\rm proj}=20$~kpc, the
spread in line-of-sight distances is expected to be about $\pm20$~kpc 
(a spread in apparent magnitude of $\pm 0.05$~dex),
which corresponds to a spread in \fehp\ of approximately $\pm 0.03$~dex.  If the
cold component is 
systematically more (less) distant than M31's spheroid (\S\,\ref{sec:gss}), the intrinsic difference
in metallicity between the populations will be slightly smaller (greater).

Although fields f115 and f135 do not show clear evidence of substructure
(\S\,\ref{sec:anal_ind}), they
are both within the minor axis range of the $\sim-300$~\kms\ cold component
(Fig.~\ref{fig:spatdist}; \S\,\ref{sec:spatdist}).
If the cold components
in fields H11, f116, and f123 have the same physical origin, it is reasonable 
to postulate that there may be substructure in fields f115 and f135 that is 
not detected by the fits
to the radial velocity distributions.  The stars in fields f115 and f135 
that fall within the triangular region marked in the upper panel of
Figure~\ref{fig:spatdist} have mean (median) \fehp\ values of $-0.67$ ($-0.57$) 
and $-0.56$ ($-0.46$), respectively. 
Stars from fields f115 and f135
that have velocities both outside the triangular region and 
$v_{\rm hel}<-150$~\kms\ have mean (median) \fehp\ values of $-0.61$ ($-0.53$) and $-0.64$ ($-0.59$), respectively.
The \fehp\ distributions of stars in fields f135 and f115 that are within the 
triangular region of the $\sim -300$~\kms\ cold component
are consistent with both the $v_{\rm inner}$ and $v_{\rm outer}$ distributions.
The difference between the substructure and spheroid metallicity distributions
is small, and is only statistically significant when the three 
fields contributing to the $\sim-300$~\kms\ cold component are combined (into
the $v_{\rm inner}$ and 
$v_{\rm outer}$ samples).  The number of stars within a restricted 
velocity range in any given field is too small to support a statistically 
significant comparison.

\subsection{Comparison to Previous Observations}
The H11 field has been presented in previous papers as a smooth spheroid field 
that is well-described by a single, kinematically hot component 
\citep{bro06,bro06b,kal06a}.  There are two related factors that have caused 
this field to be reinterpreted as containing substructure.
First, the data set presented in this paper represents an order of 
magnitude increase in the sample size of confirmed M31 RGB stars over 
previous spectroscopic samples published by our group in the SE minor 
axis region of the inner spheroid ($R_{\rm proj}\lesssim 30$).  
Only the H11 and a0 fields have 
been previously published, and the data published in H11 contained 
only a fraction of the full data set for that field, based on preliminary reductions 
(\S\,\ref{sec:dataredux}).   This paper presents the full H11 data set and 
combines it with new data from neighboring fields.  Second, the 
triangular shape of the $\sim-300$~\kms\ cold component in a plot of 
velocity vs.\ position along the minor axis implies that the debris 
has a relatively 
large velocity dispersion in the H11 field.  This makes the debris 
harder to detect against the broad underlying spheroid than if it were 
kinematically colder.  The increase in the sample size of stars with 
recovered velocities in 
the H11 field, coupled with the context provided by the velocity 
distributions in neighboring fields, has proved to be crucial in 
detecting the 
substructure in H11.

Previous observational studies have suggested the possibility of substructure 
along M31's southeastern minor axis. \citet{rei02} found 
a dynamically cold grouping of 4 metal-rich M31 RGB stars (out of $\sim 35$) 
along a southeastern minor axis field at $R_{\rm proj}\sim 19$~kpc.  This hint of 
substructure was strengthened by subsequent observations at 7 and 11~kpc along
the southeastern minor axis, which increased the total M31 RGB sample to $\sim 100$ stars 
\citep{guh02}.  The starcount maps in \citet{fer02} (their Fig.~2) also show 
hints of a population of metal-rich stars along the southeastern minor axis, and
deep HST/ACS imaging has discovered a significant intermediate-age population
in field H11 \citep[\S\,\ref{sec:intm_age},][]{bro03}.
With the spectroscopic sample of $>1000$ stars presented in this paper, we are able to confidently 
identify a cold component
along the southeastern minor axis and characterize its properties.

\section{Physical Origin of the Cold Component}\label{sec:origin}

The observed M31 RGB population along the southeastern minor axis exhibits a 
spatially
varying kinematically cold component, which has a higher mean
metallicity than the underlying inner spheroid population.  
A cold component with these properties could be part of the 
outskirts of M31's disrupted disk or debris left by disrupted satellites.  
The models of F07 predict debris stripped from the progenitor of the 
GSS should be present in these fields.  (It is also
possible, of course, that the observed substructure is satellite debris {\em unrelated} to
the GSS.) 
This section examines both the continuation of the GSS (\S\,\ref{sec:gss}) and M31's disturbed
disk (\S\,\ref{sec:disk}) as possible physical origins of the cold component. 

\subsection{Relation to the Giant Southern Stream}\label{sec:gss}
\subsubsection{Model of a Recent Interaction}
Debris in the form of coherent shells has been observed in many elliptical
galaxies; these shells are believed to be formed by the tidal disruption of a satellite
galaxy on a nearly radial orbit \citep[e.g.,][]{sch80,her88,bar92}. 
F07 presents the hypothesis that the Northeast and Western ``shelves'' 
observed in M31 (Fig.~\ref{fig:roadmap}, also see Fig.~1 of F07) are a similar phenomenon to these shell systems,
and have been created by the disruption of the progenitor of the GSS.  
Shells have coherent velocities, and display a distinctive 
triangular shape in the $v_{\rm los}$ vs. $R_{\rm proj}$ plane \citep[][F07]{mer98}.  In 
general, as $R_{\rm proj}$ approaches the boundary of the shell the spread in velocities
approaches zero, with the mean velocity at the tip of the triangle expected to 
be at the systemic velocity of the system. 

F07 present an $N$-body simulation of an accreting dwarf satellite within 
M31's potential. The 
simulations use a static bulge+disk+halo model
which is based on the 
M31 mass models in \citet{gee06} combined with the 
observed stellar density distribution in the halo \citep{guh05},
 and assumes an 
isotropic velocity distribution in the outer halo. 
The simulated satellite's physical 
and orbital properties have been chosen 
to reproduce the observed properties of the GSS and Northeast shelf, using
the methods of \citet{far06}.  
The simulations show that the orbit which reproduces these features
also reproduces a photometric feature identified in F07 as the ``Western shelf'' 
and an observed 
stream of counter-rotating planetary nebulae \citep{mer03,mer06}.   

\begin{figure}
\plotone{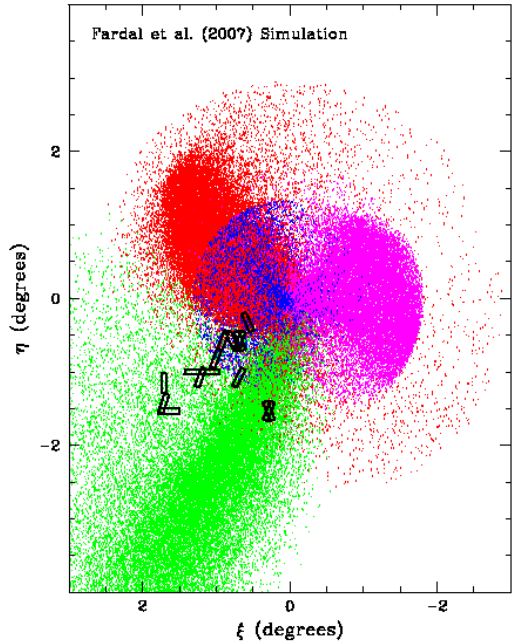}
\caption{
Projected sky position (in M31 centric coordinates $\xi$ and $\eta$)
of tidal debris in the F07 simulations of the merger of a 
dwarf galaxy with M31.  
Particles approaching their first pericentric passage are part of
the GSS (green).  Particles approaching their second pericentric passage form the Northeast shelf
(red), and particles approaching their third pericentric passage form the Western shelf
(magenta) identified in F07.  Particles in blue are approaching their fourth
pericentric passage, and form a faint shelf feature which is predicted to be
most easily visible in the southeast.
The position of our spectroscopic masks 
are also shown; fields f123 and f135 straddle the edge of the Southeast shelf (Fig.~\ref{fig:cfht}). 
 The
two masks at $\xi= 0.3$\degree, $\eta= -1.5$\degree\ are in field H13s,
which is discussed in \S\,\ref{sec:f135} and \S\,\ref{sec:intm_age}.  
}
\label{fig:xieta_sim}
\end{figure}

Figure~\ref{fig:xieta_sim} shows the projected sky positions 
(in M31-centric coordinates) of the satellite
particles from the F07 simulation.  The particles are color-coded by shell, 
or equivalently, by the number of pericentric passages
they have made.  Green particles represent particles approaching their first 
pericentric
passage; they correspond to the observed GSS.  Red particles correspond to the 
observed Northeast shelf; they are between their first and second pericentric passages.  Magenta 
particles represent the Western shelf identified in F07, and are between their second and third pericentric passages.
The blue particles are between their third and fourth 
pericentric
passages, and represent the  ``Southeast shelf'' predicted by F07.  
This last feature is predicted to extend out to a radius of 18 kpc and is expected to be very faint, 
as it consists of particles further forward in the continuation of the stream than the 
more visible Northeast and Western shelves.
This feature 
actually covers $\sim 180$\degree\ in position angle on the east side of M31,
although it is likely to only be
visible in
the southeast 
due to its overlap with the Northeast shelf and M31's disk.  
In the F07 simulations, the Northeast shelf is made up of both the leading material from the progenitor's 
first pericentric passage and trailing material from its second 
pericentric passage, while the Western shelf is formed by leading material.
The simulations are unable to constrain whether or not the 
satellite disrupts completely, as this is dependent on the central 
density of the satellite.

\begin{figure}
\epsscale{0.9}
\plotone{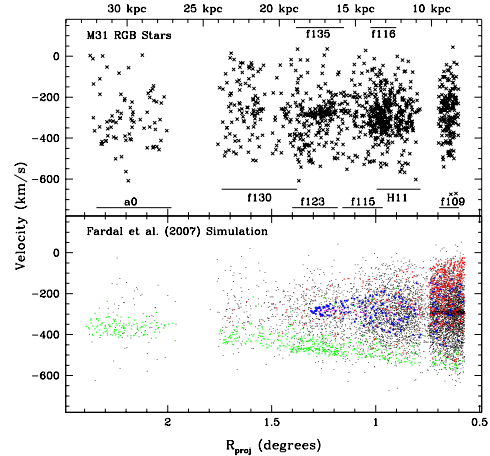}
\caption{
Line-of-sight velocity vs. projected radial distance from 
M31's center ($R_{\rm proj}$) of spectroscopically confirmed M31 RGB stars 
(\textit{top panel}) and particles from the F07 simulations
of the orbit of the progenitor of the GSS (\textit{bottom panel}).  
Particles are drawn from the locations of each of our DEIMOS 
masks. The area from which particles have
been drawn has been increased relative to the size of a slitmask to increase the
number of particles.  The 
satellite particles are color-coded according to which shell they are in: the giant
southern stream (green), the Northeast shelf (red), and the predicted Southeast shelf (blue).  Black points are particles from the bulge+disk+stellar-halo 
of M31. The blue particles form a triangular shape, with an
increasingly wide kinematic profile as the minor axis distance to the 
center of M31 decreases, as seen in the data. 
The tip of the triangle at $R_{\rm proj}\sim 1.3$\degree\ (18 kpc) in the 
simulated data agrees well with the observed tip in the data in field f123. 
The cold concentration of M31 particles at 
$v_{\rm M31}\approx 0$~\kms\ extending from $R_{\rm proj}\sim 0.58$\degree\ to $\sim 0.73$\degree\
(7.9 to 10.0 kpc) corresponds to the disk of M31 (\textit{bottom panel});  
the kinematical signature of a smooth, cold disk is not seen in our data (\textit{top panel}).  }
\label{fig:spatdist_sim}
\end{figure}

The bottom panel of Figure~\ref{fig:spatdist_sim} presents the distribution of particles from 
the F07 simulation in the $v_{\rm los}$ vs. $R_{\rm proj}$ plane.
  The figure shows particles related
to the merging satellite as well as particles associated with the static
bulge+disk+stellar-halo M31 model used in F07.  
In order to carry out a precise comparison to our observational data set, the 
F07 simulation particles were 
selected based on their projected sky position; all particles that fall inside a 
16\arcmin$\times$10\arcmin\ area (the approximate area of one DEIMOS mask is 
16\arcmin$\times$4\arcmin) centered 
on the position of our observed fields, and oriented at the position angle of our 
observed masks, are displayed.
The satellite particles are color-coded
by shell (or, equivalently, the number of orbits they have made) as in
Figure~\ref{fig:xieta_sim}.  Green particles are associated with the GSS,  
red particles with the Northeast shelf, and black particles with the bulge+disk+stellar-halo model for M31.  The blue particles, which
are part of the predicted ``Southeast'' shelf, 
form the distinctive triangular shape expected of a shell feature in the
$R_{\rm proj}$-$v_{\rm los}$ plane.  

In the following discussion, it is important to distinguish between ``spillover'' 
from the GSS (material associated
with the GSS that is in roughly the same orbital phase as the material
in the GSS, i.e., green particles in Figs.~\ref{fig:xieta_sim} \& 
\ref{fig:spatdist_sim}) versus wrapped around portions of the GSS 
(material associated
with the GSS progenitor that is leading the GSS and has undergone one or more
additional pericentric passages, i.e., red, magenta and blue particles in 
Figs.~\ref{fig:xieta_sim} \& 
\ref{fig:spatdist_sim}).  The latter is the main theme of this paper, although
we also briefly discuss the former in \S\S\,7.1.2\,--\,7.1.3.

\subsubsection{Comparison of Model to Data}

\subsubsection*{Sky Position}  
Figure~\ref{fig:xieta_sim} shows the projected sky positions (in M31-centric coordinates) of the satellite
particles from the F07 simulation as well as the
size, position and orientation of our Keck/DEIMOS slitmasks  
(\textit{rectangles}).
Fields f123 and f135 land on the edge of the predicted Southeast shelf (\textit{blue 
particles}), fields f115, f116, H11 and f109 all lie within the boundary of the Southeast shelf, and fields f130 and a0 lie beyond it.
Thus, the model predicts that the edge of the shell feature should pass
directly through our CFHT/MegaCam image.  Indeed, there is
an apparent edge visible in the CFHT starcount map (Fig.~\ref{fig:cfht},
passing through field f123), in the same location as that
predicted for the Southeast shelf.
A close inspection of the \citet{iba05} starcount map 
(Fig.~\ref{fig:roadmap}) reveals a point of bifurcation between the edge of 
the Northeast shelf and a fainter feature at $\xi\approx 1.6$\degree, $\eta\approx 0.2$\degree,
in rough agreement with the bifurcation of the two features in 
Figure~\ref{fig:xieta_sim}.  This bifurcation is more evident in the Sobel-filtered
map in F07 (their Fig. 1).  The radii of the shells in the simulation are 
robust (\S\,4.2 of F07); thus the agreement 
between the observations and the simulations is a strong confirmation of the 
validity of the F07 model.

\subsubsection*{Kinematic Trends}
The $\sim-300$~\kms\ cold component
observed in our
minor axis fields shows the distinctive triangular velocity pattern expected
of a shell feature in the $R_{\rm proj}$-$v_{\rm los}$ plane.  
Figure~\ref{fig:spatdist_sim} compares our data (top panel) to the F07 model (bottom panel).
  The distribution of observed velocities narrows to
a tip at $R_{\rm proj}\approx-1.3$\degree\ (18~kpc) in the simulated particle
distribution, which is similar to the position of the tip of the velocity
distribution in our observed data.  
At $R_{\rm proj}\approx-1$\degree\ (13.7~kpc), the velocity
distribution of the observed substructure has widened to a spread of 
$\sim200$~\kms\ (measured from the edges of the feature), 
also in agreement with the velocity spread of the predicted Southeast shelf. 

The ``boxy'' shape of the velocity distribution in field H11 
(Fig.~\ref{fig:velhist}) is also consistent with the interpretation of 
the substructure as being part of a shell system.  The velocity distributions
of shells have a clearly defined minimum and maximum line-of-sight velocity at a
given $R_{\rm proj}$.  Stars tend to congregate at the minimum and maximum
velocities \citep[][F07]{mer98}, although their location in the 
$R_{\rm proj}$-$v_{\rm los}$ plane depends on the region they occupy 
in space (cf. the discussion in F07). 
 
A maximum-likelihood Gaussian fit to the particles identified with the 
Southeast shelf and within the minor axis distance spanned by field 
f123 yields parameters of
$\langle v\rangle=-280.6$~\kms\ and $\sigma_{v}=19.4$~\kms.  A Gaussian fit to 
the Southeast shelf particles within the minor axis range spanned by fields 
f116 and H11 returns $\langle v\rangle=-292.6$~\kms\ and $\sigma_{v}=60.5$~\kms.  
The
mean velocity and dispersion of the predicted shelf is in good agreement
with the properties of the observed substructure (Table~\ref{table:rvfits}).

\subsubsection*{Metallicity Distribution}

If the substructure
identified in this paper is part of the predicted Southeast shelf in
F07, it should have a similar metallicity distribution to that of the
GSS, since the two structures originated from the same progenitor.
As part of our Keck/DEIMOS survey of M31's inner spheroid and halo, we
have taken spectra in two fields located on the GSS: a field at 
$R_{\rm proj}=33$~kpc \citep{guh06} and a field at $R_{\rm proj}=21$~kpc 
\citep[H13s;][]{kal06a}.  The cumulative \fehp\ distribution of stars identified 
kinematically as belonging to the GSS in these two fields is
plotted in the bottom panel of Figure~\ref{fig:met} (thin dotted line).
It is very similar to the \fehp\ distribution of stars that are kinematically
associated with the substructure in fields H11, f116, and f123.  
The mean and median \fehp\ of the stars in the GSS are 0.1 and 0.05 dex 
more metal-poor, respectively, than the mean and median \fehp\ of the 
substructure in fields H11, f116, and f123, after correcting for 
spheroid contamination 
(\S\,\ref{sec:met}).  The estimated 
number of inner spheroid star contaminants in the GSS sample is a few stars 
\citep{guh06,kal06a}.  If the 3 most metal-poor stars are removed 
(\fehp$<-2.25$) from the GSS distribution, the mean and median
metallicity of the GSS stars are only 0.01 dex more metal-poor than the 
the minor axis substructure.
The \fehp\ values of the GSS stars
have not been corrected for the GSS' measured distance relative to M31 
\citep[$\sim 
50$~kpc behind M31 for these 2 fields;][]{mccon03}.  Accounting for this effect
would decrease the average metallicity of the GSS by $\sim 0.1$~dex.  The
distance to the minor axis substructure is not known, although the F07 simulations predict
that the Southeast shelf should be approximately at M31's distance, with a spread in 
distances of $\pm 9.2$~kpc (this corresponds to $\pm 2\sigma$ in terms of the 
distribution of particle distances). 

\subsubsection*{Strength of the Cold Component}
The cold component comprises
44\% of the total population of observed stars in fields H11 and f116 and 31\% of
observed M31 RGB stars in field f123 (Table~\ref{table:rvfits}).  This
corresponds to a lower limit for the total fraction of stars in the cold 
component of 21.7\% in the fields within the predicted range of the Southeast 
shelf (f109, H11, f116, f115, f123, and f135).  The Southeast shelf in the 
simulations is much weaker, comprising only
2.7\% of the total population in these fields (this number increases to 3.4\%
if the number of shelf particles is compared only to the number of 
bulge+disk+stellar-halo M31 particles).  The strength of the feature in the 
simulations is highly dependent on the mass of the progenitor and the time 
since the first collision (F07).  Thus, the strength of the observed 
substructure will place interesting constraints on future models of the stream,
but cannot be used as a reliable discriminant of the applicability of
the model at the present time. 

\subsubsection*{Fields Without Clear Detection of Substructure}
We do not
find a clear detection of substructure in fields f130 and a0.  In the
context of the Southeast shelf, this is not surprising as both these
fields are beyond the radial range spanned by the shelf 
(Fig.~\ref{fig:xieta_sim}).  Field f109 is significantly inward of 
the innermost field in which we detect substructure.  In the 
simulation, the particles associated 
with the Southeast shelf continue into the region covered by field f109 with a 
spread in velocities of $\sim350$~\kms\ (Fig.~\ref{fig:spatdist_sim}, bottom panel). 
The data in field f109 is consistent with being drawn from a single Gaussian
(\S\,\ref{sec:anal_ind}) and shows no evidence of substructure.
A secondary component
with a spread in velocities as wide as predicted would be very difficult to 
differentiate from the broad spheroidal component, and would require a 
much larger sample of M31 RGB stars in this field than is currently available. 

Field f115 is well within the boundaries of the Southeast shelf.  As 
discussed in \S\,\ref{sec:spatdist}, its velocity distribution is consistent
with being drawn from the same parent distribution as field f116.   
The shell in field f115 may be difficult to detect in our data due to
the broad ($\sim 55$~\kms) nature of the substructure and the smaller 
number of stars available in this field ($\sim 50\%$ less than in 
field f116), or the shelf may be inherently clumpy.   

Field f135 is on the edge of the simulated shelf, and shows evidence
for a peak of stars near $v_{\rm hel}\sim -300$~\kms\ in its radial velocity 
histogram. 
In light of the simulations, we discuss this field in detail in 
\S\,\ref{sec:f135}, and show that it may contain a kinematically-cold 
component whose properties are consistent with both the observations 
of the $\sim-300$~\kms\ cold component
in fields H11, f116, and
f123 and the simulations of the Southeast shelf.

\subsubsection*{Debris from the GSS}
The GSS is observed 
to have an asymmetric shape, with a sharp edge on the
eastern side and a more gradual decline in density on the western side 
\citep{mccon03}.  However, the eastern edge is not an absolute one in 
the models, and ``spillover'' material from the GSS (\S\,\ref{sec:gss}) 
is predicted by the F07 simulations to be present in all of our fields.
Field f135 is the closest of our fields to the GSS' eastern edge
(Figs.~\ref{fig:roadmap} and \ref{fig:xieta_sim}); 
\S\,\ref{sec:f135} discusses the evidence for spillover debris from the 
GSS in this field. 
 
Since the density of M31 RGB spheroid stars falls off strongly with 
increasing radius in the inner spheroid, 
the contrast of cold GSS debris against the dynamically hot spheroid is
expected to be greatest in our outermost field, a0.
The GSS debris in field a0 is predicted to have a mean velocity of 
$\langle v\rangle_{\rm GSS}=-364$~\kms.  Although
field a0 shows hints of peaks in the radial velocity distribution 
at $v_{\rm hel}<-300$~\kms\
(Fig.~\ref{fig:velhist}), only a handful of these stars are as metal-rich 
as the GSS ([Fe/H]$\gtrsim -0.75$).  
 This allows us to place an upper limit on the 
contamination of field a0 by the GSS of $\lesssim 5$ stars 
($\lesssim 6\%$).   
  
Although the F07 simulations reproduce many of the observations
in the GSS, they predict a much larger amount of debris 
on the eastern side than is observed in our fields 
(Fig.~\ref{fig:spatdist_sim}).  Many factors can influence the 
structure of the debris
in the simulations, including the shape and rotation of the progenitor.
The current models of the stream (F07) use a spherical, non-rotating
progenitor.  A more complex model of the progenitor may be required to 
reproduce the observations (Fardal et al., private communication).

\subsubsection{Evidence of the Southeast Shelf and Spillover Debris from the GSS in Field f135}\label{sec:f135}

The F07 simulations predict that in addition to the Southeast shelf 
some spillover debris associated with the GSS (\S\,\ref{sec:gss}) should be present in field 
f135.  
The sky coordinates of 
field f135 are $\xi=0.7$, $\eta=-1.1$, which places it on the edge of 
both the GSS and the Southeast shelf in the F07 simulations (Figure~\ref{fig:xieta_sim}).  
In addition, the radial velocity distribution of stars in field f135 
is not well-fit by either a single or double Gaussian 
(\S\,\ref{sec:anal_ind}) and 
shows evidence of a metal-rich population (\S\,\ref{sec:met}). 
Motivated by the close match between the observations and simulations
of substructure
in fields H11, f116 and f123, 
we carry out a constrained fit of the radial velocity 
histogram of field f135 to the sum of three Gaussians to determine if the 
Southeast shelf is present in this field.  
 
\begin{figure}
\epsscale{0.85}
\plotone{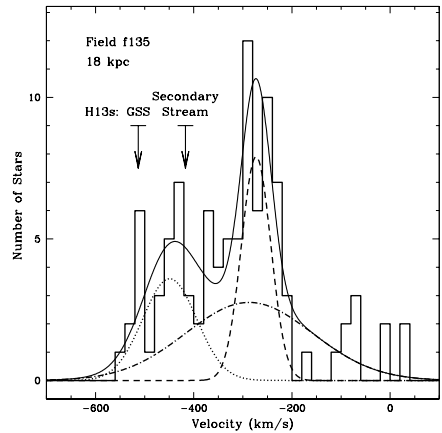}
\caption{
Radial velocity histogram of M31 RGB stars in field f135.  
A constrained triple Gaussian (\textit{solid curve}) has been 
fit to the observed
data using a maximum-likelihood technique, with rough constraints imposed on
the parameters based on the properties of the simulated substructure 
(\S\,\ref{sec:f135}).  
 The observed velocity distribution is well fit by a sum of three Gaussians: 
(i) \Gsph, the wide Gaussian which corresponds to the underlying inner 
spheroid of M31 (\textit{dot-dashed curve}, \S\,\ref{sec:anal_comb}), 
(ii) a component centered at \mvsb$=-273$~\kms\ 
with a width of \sigvsb$=30$~\kms, which comprises 30\% of 
the total population and which likely corresponds to
the Southeast shelf (\textit{dashed curve}), and (iii) a narrow component
centered at \mv$=-449$~\kms\ with a width of $\sigma_v=55$~\kms, which 
comprises 25\% of the total population 
(\textit{dotted curve}).
  The mean velocity and velocity dispersion ($\pm 1\sigma_v$) of the cold components in field H13s, at a 
similar radial distance along the GSS as field f135, are shown as arrows and
horizontal lines \citep[\S\,\ref{sec:f135};][]{kal06a}.
}
\label{fig:velhist_f135}
\end{figure}

The mean velocity and velocity dispersion of the simulated Southeast
shelf and GSS particles in 
field f135 are $\langle v\rangle_{\rm SE}=-286$~\kms, 
$\sigma^{\rm SE}_v=19$~\kms\ and $\langle v\rangle_{\rm GSS}=-458$~\kms, 
$\sigma^{\rm GSS}_v=40$~\kms.  These values were used as rough constraints for
the triple Gaussian fit: the means were allowed to vary within $\pm100$~\kms\ of the 
predicted values and the dispersions were allowed to vary  
from 1~\kms\ to $3\sigma^{\rm SE}_v$ and $2\sigma^{\rm GSS}_v$.  
The wide Gaussian component parameters were held fixed at the values for \Gsph:
\mvsph$=-287.2$~\kms\ and \sigvsph$=128.9$~\kms\ 
(\S\,\ref{sec:anal_comb}).  The maximum-likelihood triple Gaussian fit 
is displayed in Figure~\ref{fig:velhist_f135} (\textit{solid curve}).  
The wide underlying inner spheroid component (\Gsph, \textit{dot-dashed curve}) 
comprises 45\% of the population.  The Southeast shelf component 
(\textit{dashed curve}), which is the narrow peak at
\mvsb$=-273$~\kms, comprises 30\% of the population and has 
a width of \sigvsb$=30$~\kms. The ``GSS'' component (\textit{dotted curve}) at \mv$=-449$~\kms\ has 
a dispersion of $\sigma_v=55$~\kms\ and comprises 25\% of the total population. 
If a more constrained fit is carried out with the $\langle v\rangle$ and $\sigma_v$ parameters for
all three Gaussian components held fixed (at the predicted values 
for the simulated shelf and stream particles and at the parameters of the Gaussain \Gsph) 
and only the fractions of stars in the
various components are allowed to vary, the best-fit distribution has 
$N_{\rm shelf}/N_{\rm tot}=0.18$ and $N_{\rm GSS}/N_{\rm tot}=0.17$.

\begin{figure}
\plotone{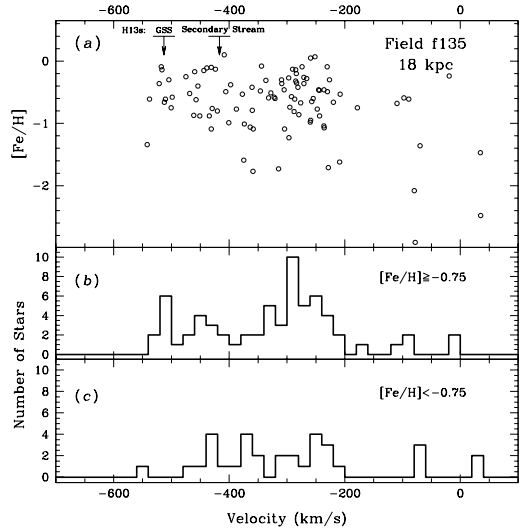}
\caption{
(\textit{a}) Metallicity vs. heliocentric velocity for the M31
RGB stars in field f135, which lies to the east of the edge of
the GSS.  The mean velocities and dispersions ($\pm 1\sigma_v$) of 
the GSS and the secondary stream from
the GSS field H13s \citep{kal06a} are marked, as in Figure~\ref{fig:velhist_f135}.
(\textit{b}) Velocity distribution of stars with \fehp$>-0.75$.  In addition
to the $\sim-300$~\kms\ cold component, the metal-rich subset
shows evidence of concentrations of stars corresponding to the GSS and secondary
stream in the H13s field. 
(\textit{c}) Velocity distribution of stars with \fehp$<-0.75$. The 
velocities of stars with
metallicities lower than $-0.75$~dex appear evenly distributed.
}
\label{fig:f135_met_vel}
\end{figure}

The kinematic properties (mean velocity and 
velocity dispersion) of the Southeast shelf component in the triple 
Gaussian fit to field f135 are consistent not only 
with the simulations, but also with 
what one would expect for the Southeast shelf in this field based on the 
observations (e.g., Fig.~\ref{fig:spatdist_sim}, 
Table~\ref{table:rvfits}).  In addition, the fraction of the population
in f135 which is in this component is consistent with the fraction of the
population which is identified with the Southeast shelf in fields 
H11, f116, and f123.
As further evidence that the Southeast shelf is detected in field f135, 
Figure~\ref{fig:f135_met_vel} shows \fehp\ vs. $v_{\rm hel}$ for the M31
RGB stars in field f135 (panel \textit{a}), as well as velocity histograms
for stars in two \fehp\ bins [(\textit{b}) \fehp$>-0.75$ and (\textit{c})  
\fehp$<-0.75$].   
As in fields H11, f116 and f123, the substructure in field f135  
that is identified with the Southeast shelf (\mv $=-273$\kms) is 
relatively metal-rich.  
 
The velocity dispersion of $\sigma_v=55$~\kms\ for the ``GSS component'' 
inferred from the first
of the triple-Gaussian fits above is large compared
to previous measurements of $\sim 15$~\kms\ for the dispersion of the GSS 
\citep{iba04,guh06,kal06a}.  In addition, although the best-fit \mv$^{\rm sub}$
of the most negative component in f135 is similar to that of the F07 model,
the predicted GSS mean velocities in the F07 model are not as negative as the 
observed velocities of the GSS.   \citet{kal06a} analyzed a field centered 
on a high surface brightness protion of the GSS, at approximately the 
same radial distance along the stream as field f135 (field H13s, located at 
$\xi=0.3$\degree, $\eta=-1.5$\degree\ in Figure~\ref{fig:xieta_sim}) 
and found a 
secondary cold component, the ``H13s secondary stream,'' whose origin 
and physical extent are unknown.  
The observed mean velocities of the GSS and secondary stream in the 
H13s field are $\langle v \rangle^{\rm GSS}=-513$~\kms\ and 
$\langle v \rangle^{\rm sec.str.}=-417$~\kms, respectively,
with velocity dispersions of $\sigma_v^{\rm GSS}=\sigma_v^{\rm sec.str.}=16$~\kms\ \citep{kal06a}.
The mean velocities 
and velocity dispersions ($\pm 1\sigma_v$) of these two components are 
shown as arrows with horizontal lines
in Figures~\ref{fig:velhist_f135} and \ref{fig:f135_met_vel}.  
There appear to be two metal-rich peaks in f135 with the approximate velocities 
of the GSS and secondary stream in H13s.
If both the GSS and H13s secondary stream 
are present in field f135, they appear
in approximately equal proportion.
In H13s
the GSS dominates over the secondary stream by a factor of two \citep{kal06a}. 

In the simulations, the GSS dominates over the Southeast shelf in 
field f135 by about a factor of 10; the observations indicate that, at best,
these two populations are roughly equal.  The simulations predict that the GSS
should comprise a total of 35\% of the population in field f135, which
is somewhat larger than the fraction of the total population in the
most negative cold component from the triple-Gaussian fit shown in 
Figure~\ref{fig:velhist_f135} (25\%).  If the ``GSS''
component from the triple-Gaussian fit
is actually comprised of two narrower streams,
the GSS comprises a much smaller percentage of the stars in this field than
predicted by the simulations.

\subsection{Arguments Against a Disk Origin}\label{sec:disk}

%

The substructure discovered in fields H11, f116 and f123 is centered at close to the systemic velocity
of M31 ($v_{\rm sys}=-300$~\kms), which is also the radial velocity expected
for an M31 disk component on the minor axis. 
Recent observational evidence suggests that M31's stellar disk 
extends smoothly out to $R_{\rm disk}\sim 40$ kpc and has a velocity dispersion 
of $\sim30$~\kms; isolated features with disk-like 
kinematics have been observed as far out as $R_{\rm disk}\sim 70$ kpc 
\citep{rei04,iba05}.  We present three lines of evidence
arguing against a disk origin for the $\sim-300$~\kms\ cold component found
on the southeastern minor axis.

\subsubsection*{Disk to Inner Spheroid Surface Brightness Ratio}
Previous 
measurements of M31's 
stellar disk and inner spheroid \citep{wal88,pri94} indicate that the 
disk constitutes a negligible fraction of the 
total light in our minor axis fields.  Even in our innermost field f109 
(at $R_{\rm proj}\sim 9$~kpc, 
corresponding to $R_{\rm disk}\sim 38$~kpc for a disk inclination of 77\degree), 
the disk fraction is expected
to be only $\sim10\%$ \citep{guh05} based on disk scale radii of 5.0\,--\,6.0~kpc
\citep{iba05}.  We see no evidence of a cold, disk-like feature in f109's 
radial velocity distribution, although a $\lesssim10$\% component could 
be difficult to detect.
The disk fraction drops sharply at larger radii: for example, at the 
distance of our innermost field containing the $\sim-300$~\kms\ 
cold component, H11 ($R_{\rm disk}\sim 51$~kpc), the expected smooth 
disk fraction is $\sim1\%$ \citep{bro06}.

Non-uniformities in M31's stellar disk could 
result in a higher disk fraction in our fields.  To explain the 
strength of the $\sim-300$~\kms\
cold component in field H11 ($N_{\rm sub}/N_{\rm tot}=44$\%), the disk would have to contain a $45\times$ 
enhancement in this field relative to the smooth disk.  Even more extreme disk 
enhancements are needed to explain the cold component in fields f116 and f123.

If there is a warp such that the outer disk is more face on than the inner 
disk (\citet{iba05} find a best fit disk inclination angle of 64.7\degree\
from $20<R_{\rm disk}<40$~kpc), the effective disk radii of our fields will
be smaller and the smooth disk contribution larger than in the above
calculation.  As an extreme example of a warp, we consider a disk of scale
length 5.7~kpc whose inclination changes from 77\degree\ at small radii
($R_{\rm disk}<20$~kpc) to 60\degree\ at large radii ($R_{\rm disk}>20$~kpc).
In this case, H11 is at $R_{\rm disk}\sim 35$~kpc while f123 is at 
$R_{\rm disk}\sim 47$~kpc.  With this disk model, taking into account 
projection effects in the disk surface brightness (the surface brightness
decreases as the disk becomes more face on) but ignoring dust effects, 
the expected disk fractions in fields f109, H11, f116, and f123 are 20\%,
18\%, 17\%, and 10\%, respectively.  Thus, it would still require a
$\sim 2.5\times$ enhancement in the disk to explain the cold component in
H11.  More importantly, the 20\% cold, smooth disk fraction predicted by
this warp model in field f109 is inconsistent with our radial velocity
data (Fig.~\ref{fig:velhist}).  This disk fraction estimate for the warp is
optimistic (high) in that it does not account for the reduction in surface
brightness that would be caused by any stretching associated with the
putative warp.

 
\subsubsection*{Velocity Dispersion}
The measured velocity dispersion of the $\sim-300$~\kms\
cold component in fields H11 and f116 is $\gtrsim50$~\kms, while the velocity
dispersion of this component is only $\sim10$~\kms\ in field f123.  These
measurements are significantly above and below, respectively, the typical
velocity dispersion of $30$~\kms\ measured for the extended, disk-like
structure by \citet{iba05}, for which they observe a range in velocity
dispersions from $\sim 20$\,--\,$40$~\kms.  If the 
cold component in each of these three fields is from the disk, it would require
a warp and non-uniformities in the disk that happen to have the observed 
triangular shape in the position-velocity plane (top panel of
Figure~\ref{fig:spatdist}).  
 
\subsubsection*{Stellar Ages and Metallicities}
If the $\sim-300$~\kms\ cold component is
debris associated with M31's extended disk, its stellar population should 
reflect this.  However, a comparison
of the stellar ages and metallicities of the H11 field 
with deep HST/ACS and ground-based imaging of disk-dominated M31 
fields yields very different star formation
histories.  Obviously disturbed sections of M31's disk (e.g., the Northern 
Spur and the G1 clump) show evidence of recent star formation 
\citep[$\sim 3$~Gyr ago in the Northern Spur and $\sim 250$~Myr ago 
in the G1 clump;][]{fer05}, while deep HST/ACS imaging of field H11
reveals that very few of the stars are younger than 4 Gyr \citep{bro03,bro06b}.
\citet{bro06b} find that their disk-dominated HST/ACS field H13d, located at 25~kpc
along the northeastern major axis, contains a significantly younger and more
metal-rich population than H11.
In addition, \citet{bro06} find very good agreement between the 
stellar populations of field H11 and an HST/ACS field on the GSS (H13s, discussed 
further in \S\,\ref{sec:intm_age}).

The comparison of the stellar ages and metallicities of M31 disk fields vs.\
that of field H11 is complicated by several uncertainties.  A radial gradient
in disk properties could result in M31's outer disk (in field H11)
being more metal-poor and older than the inner parts of the disk.  Also, the 
\citet{bro06b} {\it HST/ACS\/} field H13d likely includes multiple galactic
components including spheroid and wrap-around debris from the progenitor of
the GSS, although the radial velocity distribution indicates that M31's disk
is the dominant component in this field \citep[see Fig.~9 of][]{kal06a}.
 
\bigskip
In conclusion, the observations disfavor an extended rotating disk model as 
the physical origin of the $\sim-300$~\kms\ cold component identified along
M31's southeastern minor axis.  Although a disk origin for this substructure 
cannot be ruled out, it requires simultaneous contrivance of multiple 
properties
of the disk to explain the observations: patchy disk structure (to explain
the absence of an observed disk in f109), a warp and 
enhancement (which cannot be due to recent star formation) of the disk
to explain the strength of the cold component,
an anomalously large velocity dispersion for fields H11 and f116
and a smaller than average velocity dispersion in field f123, 
and a significant radial gradient in the metallicity and age of the stellar 
disk populations.  Compared to the elegance of the southeastern shelf
interpretation based on F07's simulations, which is a true prediction and
explains the observed properties of the cold component in 
all the fields in which it is 
detected, the disk origin clearly fails the test of Occam's razor for the most 
likely physical origin of the $\sim-300$~\kms\ cold component.

\section{Implications for the Intermediate-Age Spheroid Population}\label{sec:intm_age}

\citet{bro03,bro06,bro06b} present HST/ACS photometry of fields in M31 down to 1\,--\,1.5 
magnitudes below the main-sequence turnoff.  Our field H11 
is coincident with the 
\citet{bro03} spheroid field (Fig.~\ref{fig:cfht}).  The photometry presented in \citet{bro06} 
is from a field on the GSS of Andromeda at a projected radial 
distance of 20 kpc, and is coincident 
with the Keck/DEIMOS spectroscopy field H13s presented in \citet{kal06a} 
and shown in Figure~\ref{fig:xieta_sim} ($\xi= 0.3$\degree, $\eta= -1.5$\degree).  
In the ``smooth'' spheroid field, \citet{bro03} find that 
$\sim30$\% (by mass) of the stellar population is intermediate-age (6\,--\,8 Gyr) and 
metal-rich, while another 30\% of the population is old (11\,--\,13.5 Gyr) and
metal-poor.  \citet{bro06} find remarkable agreement in the CMDs of the stream and 
spheroid fields, indicating that the two fields have very similar age and
metallicity distributions.  They query whether the similarities between the 
populations could ``be explained by the stream passing through the spheroid field,'' but
note that this explanation is problematic:
the stream would have to dominate the spheroid by approximately the same 
factor in both fields \citep[3:1 based on a kinematical analysis
of the stream field H13s;][]{kal06a}, yet the kinematical profiles of the
two fields are distinctly different, with the H11 field failing to show 
the cold ($\sigma_{\rm v}=16$~\kms) signature of the stream seen 
in the H13s field.  However, they presciently suggested that the similarity
in the two populations (spheroid and stream) implies that ``the inner
spheroid is largely polluted by material stripped from either the stream's
progenitor or similar objects.''   

In light of the substructure presented in this paper, this seems to be
the correct interpretation of the similarity between the ``spheroid'' and 
GSS stellar populations.  
The spatial and kinematic properties of the substructure suggest that the 
region of the spheroid imaged in the original HST field \citep{bro03} is 
in fact contaminated by stars from the progenitor of the GSS.  The 
kinematical signature of the substructure at the minor axis distance of 
the HST/ACS field (H11) is both predicted 
and observed to be relatively wide (Fig.~\ref{fig:spatdist_sim}), and thus 
less obvious against the underlying hot component.  In the context of 
the F07 simulations, the minor axis substructure
is not isolated, but is part of one of a series of shells caused by the 
disruption of the GSS' progenitor, which collectively contaminate a 
large part of the inner spheroid of M31 (Fig.~\ref{fig:xieta_sim}). 
 
The current analysis suggests that $\sim 45$\% of the M31 RGB stars in the 
H11 field are in fact part of the $\sim-300$~\kms\ cold component, and not part of the broad
spheroid.  In the H13s field, 75\% of the M31 RGB stars are part of a cold 
component \citep{kal06a}.
This difference in substructure fraction agrees nicely with the difference in
the fraction of intermediate-age ($<10$ Gyr), metal-rich stars found in the stream and
spheroid fields in \citet{bro06b}: 70\% vs. 40\%, respectively.  However, 
recent HST/ACS observations of a field in the location of our f130 masks
at 21~kpc imply that this is not the end of the story: \citet{bro07} find that
the stellar population in H11 can \textit{not} be fit by a linear 
combination of the GSS (H13s) and the 21~kpc spheroid (f130) stellar 
populations, due largely to the presence of a greater number of stars 
younger than 8~Gyr in H11 than in the GSS field.  
Nevertheless, the observational evidence, combined with the theoretical predictions
of F07, strongly favor the explanation that the age and metallicity
distributions of the stream and spheroid HST fields are so remarkably similar 
because the same progenitor polluted both fields with substructure.      

\section{Summary}\label{sec:concl}
The use of the diagnostic method described in \citet{gil06} has enabled us
to isolate the first sample of spectroscopically confirmed M31 RGB stars defined \textit{without} the 
use of radial velocity.  We use this 
sample of $\sim 1000$ M31 RGB stars to measure the velocity dispersion of the inner spheroid of M31; in
the radial range $R_{\rm proj}=9-30$~kpc the inner spheroid
has a velocity dispersion of 
\sigvsph=129~\kms. Our data show no evidence of a decrease in the velocity 
dispersion over this radial range. 

The stellar radial velocity distribution in these fields shows evidence of a significant amount 
of substructure.  Compared to the large velocity dispersion seen in the 
underlying hot spheroid
population, the substructure is kinematically cold, exhibiting 
a decrease in velocity dispersion
with increasing projected radius.  In the fields in which the $\sim-300$~\kms\ cold component is observed,  
$\approx41$\% of the stars are estimated to belong to it;  the rest are
members of the hot inner spheroid of M31. The metallicity of the substructure
is higher than that of the broad
spheroidal component in the fields in which it is observed.  
 
The physical origin 
of the substructure discovered in this paper is most likely tidal debris stripped from
the progenitor of the GSS.  The data agree very 
well with the location and kinematical properties of the Southeast shelf 
predicted by 
the F07 simulations of the disruption of the GSS' progenitor,
and will add
significant observational constraints to those already existing from 
the GSS,
Northeast shelf, and
Western shelf, enabling detailed modeling 
of M31's dark matter distribution (F07).  The minor axis fields 
also place constraints on the spatial distribution of the GSS itself.   
The GSS contamination in our minor axis fields is much smaller
than predicted by the current models of the stream (F07),
which suggests the stream's progenitor had a more complex structure
than the spherical, non-rotating models used so far.

The newly-discovered substructure sheds light on the discovery of a 
significant intermediate-age population in the ``smooth''
spheroid field by \citet{bro03}, and the subsequent discovery of the 
similarity in ages and metallicities of the stars in the spheroid field and 
a field on the GSS \citep{bro06, bro06b}.  The spheroid 
HST/ACS field 
was not in fact placed on a ``smooth'' spheroid field, and the 
intermediate-age population may be part of the substructure observed in
this field.  If the
substructure identified in this paper is indeed from the same progenitor as the giant southern stream,
it is not surprising that the two HST/ACS fields would have very similar age
and metallicity distributions.  Given the number of observed fields in the inner 
spheroid which are contaminated by substructure, both in the current work and
in the literature \citep{irw05,fer05,kal06a}, it seems likely that the inner
spheroid is highly contaminated by tidal debris.  A ``smooth'' inner spheroid 
field may in fact be a rarity.

\acknowledgments 
We are grateful to Sandy Faber and the DEIMOS team for building an
outstanding instrument and to Mike Rich for his role in the acquisition of
many of the Keck/DEIMOS masks.  We thank Peter Stetson, Jim Hesser, and James Clem for
help with the acquisition and reduction of CFHT/MegaCam images, Phil Choi,
Alison Coil, Geroge Helou, Drew Phillips, and Greg Wirth for observing some
DEIMOS masks on our behalf, Drew Phillips for help with
slitmask designs, 
Jeff Lewis, Bill Mason, and Matt Radovan for fabrication of slitmasks, and
the DEEP2 team for allowing us use of the {\tt spec1d}/{\tt zspec} software.
We also thank Tom Brown for stimulating discussions and comments on the draft.  The {\tt spec2d} data reduction pipeline for DEIMOS was developed at UC
Berkeley with support from NSF grant AST-0071048.  This project was supported
by an NSF Graduate Fellowship (K.M.G.), NSF grants AST-0307966 and
AST-0507483 and NASA/STScI grants GO-10265.02 and GO-10134.02 (P.G., K.M.G.,
and J.S.K.), NSF grant AST-0205969 and NASA ATP grants NAGS-13308 and NNG04GK68G (M.F.), NSF grants AST-0307842 and
AST-0307851, NASA/JPL contract 1228235, the David and Lucile Packard
Foundation, and The F.~H.~Levinson Fund of the Peninsula Community Foundation
(S.R.M., J.C.O., and R.J.P.) and NSF grant AST-0307931 (D.B.R.). 
J.S.K. is supported by NASA
through Hubble Fellowship grant HF-01185.01-A, awarded by the Space Telescope
Science Institute, which is operated by the Association of Universities for
Research in Astronomy, Incorporated, under NASA contract NAS5-26555.
%

\appendix
These appendices are intended to give the interested reader more insight into the
origin of the differences in the distributions of the M31 RGB samples
shown in Figure~\ref{fig:velhist_bias} (\S\,A.1).  The amount of 
contamination vs.\ completeness for different \ol\ thresholds is discussed in
\S\,A.2.  Finally, in \S\,A.3 we quantify the effect of 
the dwarf contamination on the radial velocity distributions for the 
M31 RGB sample used in this analysis.
 
\subsection{A.1~Effect of the Radial Velocity Diagnostic}\label{app}
As discussed in \S\,\ref{sec:sample}, empirical probability distribution
functions (PDFs)
based on training sets of M31 RGB and MW dwarf stars are used to determine the
probability that an individual star is an M31 red giant (\pg) or MW dwarf (\pd) in 
4 (5 for field a0) diagnostics.  These probabilities are used to compute     
the likelihood a star $i$ is a red giant in a given diagnostic $j$:
\begin{equation}\label{eqn:lij}
L_{ij}~=~\log\left(\frac{P_{\rm giant}}{P_{\rm dwarf}}\right).
\end{equation}
A star's overall likelihood of being an M31 RGB star is defined as
\begin{equation}
\langle{\it L}_{\it i}\rangle= \frac{\displaystyle\sum_j w_j L_{ij}}{\displaystyle\sum_j
 w_j}.
\end{equation}
All available diagnostics for a star are given a weight of one, unless the star
is an outlier with respect to both the M31 RGB and MW dwarf PDFs in a 
two-dimensional diagnostic, in which case the weight of that diagnostic is reduced \citep[\S\,3.3 of][]{gil06}.

\begin{figure}
\plotone{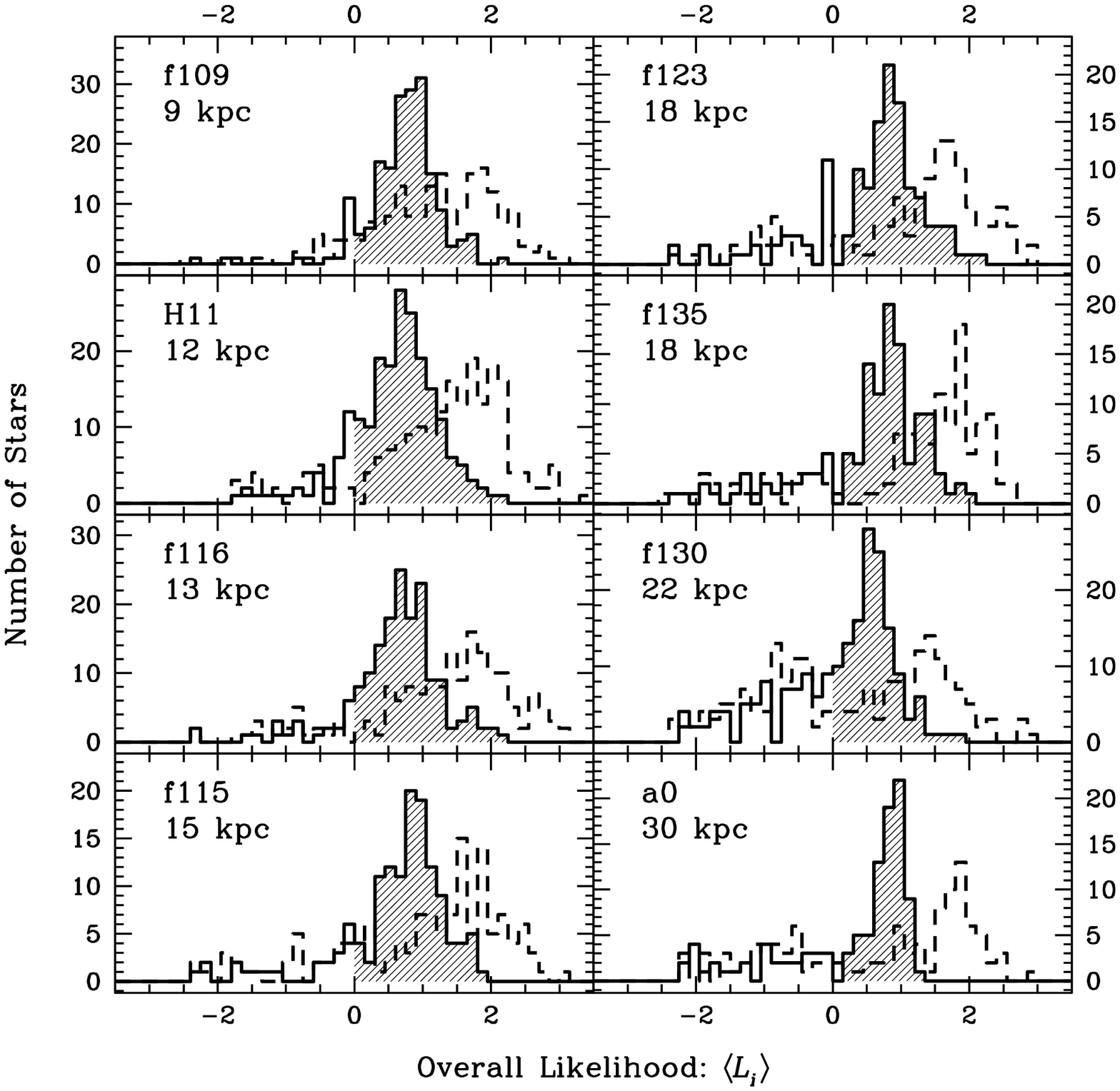}
\caption{
Overall likelihood distributions with (\ol, \textit{dashed}) and without
(\olnv, \textit{solid}) the inclusion of the radial velocity diagnostic. The analysis 
in this paper is based on the sample of M31 RGB stars shown by the shaded
histograms.  For stars with \ol$ > 0$, removal of the radial velocity 
diagnostic from the \ol\ calculation causes a shift in the peak of the 
distribution from \ol$\simeq 2$ to \ol$\simeq 1$.  This is expected, 
since radial velocity is the most
powerful of the five diagnostics.  A similar shift of the peak of the 
distribution towards \ol$=0$ likely occurs for stars with \ol$<0$, however
the number of stars in this category is too small for this shift to be
easily discernible.  
}
\label{fig:lklhd}
\end{figure}

Figure~\ref{fig:lklhd} shows the overall likelihood distributions for 
each field, both with (\ol, \textit{dashed histogram}) and without (\olnv, \textit{solid histogram}) the radial 
velocity diagnostic included in the computation.  In general, stars with
\ol\ $>0.5$ are designated secure M31 red giants and stars with \ol\ $<-0.5$
are designated secure MW dwarfs, while stars with $0<$ \ol\ $<0.5$ are designated
marginal M31 red giants and stars with $-0.5<$ \ol\ $<0$ are designated 
marginal MW dwarfs \citep[\S\,3.5 of][]{gil06}.  

Figure~\ref{fig:velhist_bias} shows the radial velocity distributions of 
several combinations of stars: only secure M31 RGB stars, secure and marginal
M31 RGB stars, and secure M31 RGB stars plus all stars that are classified
as marginal (M31 red giants and MW dwarfs), chosen by use of the diagnostic 
method both with and without the inclusion of the radial velocity diagnostic. Radial velocity 
is the most powerful single
diagnostic, with a large range of $P_{\rm giant}/P_{\rm dwarf}$ values 
\citep[e.g., Fig.~10 and \S\,3.4 of][]{gil06} and the ability to significantly boost the
combined likelihood values.  Thus, when the radial velocity diagnostic is 
not included in the overall likelihood calculation, the \olnv\ values are in general smaller, 
as can be seen in Figure~\ref{fig:lklhd}.  This causes the significant 
decrease in the number of secure M31 RGB stars seen in the top panel of
Figure~\ref{fig:velhist_bias}.  

Stars near and more negative than the 
systemic velocity of M31 have
a particularly high probability of being M31 RGB stars in the radial 
velocity diagnostic: stars with \vhel\ $=-250$~\kms\ have an $L_v=2$ 
while stars with \vhel\ $<-300$~\kms\ have $L_v=5$.  Consequently, the 
velocity diagnostic has a large affect on the \ol\ values of these stars, 
and they are 
statistically the most affected by removing 
the radial velocity diagnostic from the overall likelihood calculation.  Stars that have 
velocities in the range over which the radial velocity distributions of 
the M31 RGB and MW dwarf stars overlap ($\approx -200 < v_{\rm hel} < -125$~\kms) 
will have likelihood values near 0 in the radial velocity diagnostic.  
This means two things: (1) since the 
$L_v$ value is near zero, it adds no power to the \ol\ determination, and thus
the number of objects classifed as secure M31 RGB stars should be similar 
regardless of whether or not the radial velocity diagnostic is used, and (2) 
since these stars do not have the power of the radial velocity diagnostic, they
are statistically more likely to land in the marginal ($-0.5<$\ol $< 0.5$) regime.  The first 
effect can be seen in both the top and middle panels of Figure~\ref{fig:velhist_bias}, and the 
second effect can be seen by comparing the numbers of stars in the shaded/dotted histogram in the range 
$-200 < v_{\rm hel} < -125$~\kms\ in the top, middle, and bottom panels.  

The shifting of the overall likelihood distributions to 
smaller absolute \ol\ values due to the removal of the radial velocity diagnostic
also increases the numbers of marginal MW dwarf 
stars (and correspondingly decreases the number of secure MW dwarf stars).  As in the case of the secure M31 RGB stars, removing the radial 
velocity diagnostic from the overall likelihood calculation will have the 
largest effect on stars that have radial velocities close to or more positive
than the peak of the dwarf distribution
\citep[observed to be at \vhel$\sim -50$~\kms\ in our data set; e.g., Fig.~2 of][] 
{gil06}.  This causes stars that otherwise would be classified as secure MW dwarf stars 
to fall into the marginal ($-0.5<$ \olnv\ $<0.5$) or even secure (\olnv$>0.5$) M31 RGB regime,
depending on the likelihood values of the other diagnostics.  This  
boosts the number of
stars in this velocity range in the M31 RGB samples selected without the inclusion
of the radial velocity diagnostic. 

\subsection{A.2~Selection of the M31 RGB Sample}\label{app2}
 
Figure~\ref{fig:velhist_bias} shows that there is no difference
in the velocity distributions of secure (\ol$>0.5$) and secure plus marginal 
(\ol$>0.0$) M31 RGB stars 
with $v_{\rm hel}<-300$~\kms, chosen with the use of the radial velocity 
diagnostic.  Since
there is minimal MW dwarf contamination at velocities this negative, 
we assume that
the number of M31 RGB stars with \ol$>0$ (465 stars) represents the 
true number of M31 
RGB stars observed in this velocity range.  This allows a calculation of the 
level of incompleteness in other samples.  There are 446 stars with 
$v_{\rm hel}<-300$~\kms\ and \olnv$>0$; this sample is
96\% complete.  
Samples selected
using the thresholds \olnv$>$[0.5, 0.75, and 1] are [78\%,
57\%, and 26\%] complete.

The vast majority of the dwarf contaminants will have $v_{\rm hel}>-150$~\kms.
Counts of stars in this velocity range for a given sample, compared to 
an estimate of the expected number of M31 RGB stars in this velocity range, 
gives an
estimate of the level of dwarf contamination.
The thresholds \olnv$>0.75$ and \olnv$>1$ are quite severe, requiring
that a star have a probability of being an M31 RGB star that is 5 or 10 times
higher than its probability of being an MW dwarf star, respectively. These
samples are assumed to have minimal dwarf contamination.  Based on the numbers
of stars with $v_{\rm hel}>-150$~\kms\ that pass these \olnv\ thresholds (41
stars for \olnv$>0.75$ and 22 stars for \olnv$>1$) 
and the completeness factors calculated above, a complete sample of M31 RGB
stars would have approximately 76 stars at $v_{\rm hel}>-150$~\kms.  A sample selected using the threshold
\olnv$>0.5$ has 66 stars, which is 87\% complete.  However,
the \olnv$>0.5$ sample is expected to be only 78\% complete, implying that 9\%
of the stars with $v_{\rm hel}>-150$~\kms\ in this sample are MW dwarf star 
contaminants (overall, this sample has a 1.9\% contamination rate).  A similar
calculation for stars selected using the \olnv$>0$ threshold yields a 
contamination rate of $\sim 40$\% for stars with $v_{\rm hel}>-150$~\kms.  
This large local contamination rate corresponds to only a 5.0\% contamination rate
for the entire sample.  A parallel calculation using the number of stars with 
$v_{\rm hel}>-200$~\kms\ yields similar overall contamination rates, confirming
that the majority of the MW dwarf contamination is at $v_{\rm hel}>-150$~\kms.
The majority of the dwarf contamination comes from blue stars 
[panel (\textit{d}) of Fig.~\ref{fig:vel_met}], since the \nai,
CMD, and [Fe/H] diagnostics are all less sensitive for blue stars \citep[\S\,4.1.2 of][]{gil06}.

The focus of this paper is the substructure discovered along the southeastern minor axis
of M31. 
Therefore,
we have chosen to use the most complete M31 RGB sample chosen without the use
of radial velocity, \olnv$>0$, as it 
gives the most robust
statistics for the cold component, even though it suffers from significant 
MW dwarf contamination at velocities near zero.  The removal of the radial velocity 
diagnostic yields a sample composed of an underlying, kinematically unbiased 
M31 RGB population, as well as a contaminating MW dwarf population at 
velocities near zero.  For the analysis presented in this paper, this is 
preferable to a clean, but kinematically biased
M31 RGB sample, as the effect of the MW dwarf contamination on the
measured properties of the M31 RGB sample can be quantified by analysing
samples chosen using stricter \olnv\ thresholds.

\subsection{A.3~Bias in Measured Spheroid Velocity}\label{app3}
The main effect of the MW dwarf contamination in our chosen M31 RGB sample
(\olnv$>0$) is a positive shift in the measured mean velocity of the spheroidal 
distribution.  This
shift will affect the measured \mvsph\ in the double-Gaussian fit to 
the combined sample (\S\,\ref{sec:anal_comb}), and the \mvsph\ values 
from the single-Gaussian fits to individual fields (\S\,\ref{sec:anal_ind}). 
The substructure component is cold enough that its measured
properties are not significantly affected by the dwarf contamination.  
As discussed above (\S\,A.2), the amount of MW dwarf contamination 
is small in the \olnv$>0.5$ sample and minimal in the \olnv$>0.75$ sample.   
Although these samples suffer from significant incompleteness (22\% and 
43\%, respectively), they are 
large enough to allow a measurement of the underlying spheroidal component.   
Maximum-likelihood double-Gaussian fits to the \olnv$>0.5$ and \olnv$>0.75$ 
samples yield \mvsph\ values of $-302.5^{+7.7}_{-8.3}$ and 
$-309.6^{+10.8}_{-8.7}$~\kms, respectively.  The best-fit \sigvsph\ values for the two samples are
$123.3^{+8.3}_{-7.0}$ and $128.3^{+10.8}_{-8.7}$~\kms, respectively.  Thus,
the best-fit \mvsph\ values from the \olnv$>0$ sample should be adjusted by
$\sim 15$ to 20~\kms\ to account for the MW dwarf contamination in the sample. 
This makes the \mvsph\ value from the best-fit double Gaussian  
(\mvsph$=-287.2$~\kms) consistent with the systemic velocity of M31 ($v_{\rm sys}=-300$~\kms).
Although the overall level of dwarf contamination is expected to increase with radius due to the 
decreasing surface density of M31 stars, there is little difference in 
the mean offset of \mvsph\ if the sample is split by radius into inner 
and outer bins.


\end{document}